\documentclass{article}

\usepackage{arxiv}

\usepackage{natbib}
\usepackage{graphicx}
\usepackage{epstopdf, epsfig}
\usepackage{amssymb}
\usepackage{amsmath}
\usepackage{hyperref}
\usepackage[nameinlink, noabbrev]{cleveref}    
\usepackage{xcolor}
\usepackage{subfig}
\usepackage{float}
\usepackage[normalem]{ulem}
\usepackage{soul}
\usepackage{colortbl,xcolor}

\hypersetup{
    colorlinks=true,
    linkcolor=blue,
    citecolor=blue,
    filecolor=magenta,      
    urlcolor=cyan,
    hypertexnames=true
    }

\usepackage{booktabs}       
\usepackage{nicefrac}       
\usepackage{microtype}      

\newcommand{\bs}{\boldsymbol}
\newcommand{\bd}[1]{\mbox{\boldmath$#1$}}
\usepackage{ulem}

\newcommand{\revYJ}[1]{{\color{black}{#1}}}
\newcommand{\revJK}[1]{{\color{black}{#1}}}

\newcommand\Str{\mbox{\textrm{St}}}  
\newcommand\Rey{\mbox{\textrm{Re}}}  

\title{Influence of three-dimensionality on wake synchronization of oscillatory cylinder}

\author{Youngjae Kim$^1$\thanks{Corresponding author: youngjaekim@ucla.edu} ,
        Vedasri Godavarthi$^1$,
        Laura Victoria Rolandi$^1$,
        Joseph T. Klamo$^2$,
        Kunihiko Taira$^1$ \\
        $^1$Department of Mechanical and Aerospace Engineering, University of California, Los Angeles, CA 90095, USA \\
        $^2$Systems Engineering Department, Naval Postgraduate School, Monterey, CA 93943, USA
        }

\begin{document}

\maketitle
\begin{abstract}

We investigate the effect of three-dimensionality on the synchronization characteristics of the wake behind an oscillating circular cylinder at $\Rey = 300$. \revYJ{Cylinder oscillations in rotation, transverse translation, and streamwise translation are considered.} We utilize phase-reduction analysis, which quantifies the phase-sensitivity function of periodic flows, to examine the synchronization properties. Here, we present an ensemble-based framework for phase-reduction analysis to handle three-dimensional wakes that are not perfectly time-periodic. Based on the phase-sensitivity functions, synchronizability to three types of cylinder oscillations is evaluated. In spite of similar trends, we find that phase-sensitivity functions involving three-dimensional wakes are lower in magnitude compared to those of two-dimensional wakes, which leads to narrower conditions for synchronization to weak cylinder oscillations. We unveil that the difference between the phase-sensitivity functions of two- and three-dimensional flows is strongly correlated to the amplitude variation of the three-dimensional flow by the cylinder motions. This finding \revJK{reveals} that the \revYJ{cylinder motion modifies the three-dimensionality of the wake as well as the phase of vortex shedding, which leads to reduced phase modulation.} The synchronization conditions of three-dimensional wakes, predicted by phase-reduction analysis, \revJK{agree with} the identification by parametric studies \revJK{using} direct numerical simulations for \revYJ{forced} oscillations \revYJ{with small amplitudes}. This study presents the potential capability of phase-reduction to study synchronization characteristics of complex flows.

\end{abstract}

\section{Introduction}    \label{sec:Introduction}

Synchronization of the wake behind a bluff body to its structural vibration is of interest for various engineering applications. Such synchronization can cause detrimental effects, such as structural fatigue and resonance induced by vortex-induced vibrations \citep{sarpkaya2004critical,williamson2004vortex}. Alternatively, promoting synchronization can enhance the performance in other engineering systems, such as the molecular mixing rate in chemical reactors \citep{celik2009mixing}, the heat transfer rate of heat exchangers \citep{gau2001synchronization}, and the efficiency of energy-harvesting systems \citep{wang2020state}. The importance of accurately predicting synchronization conditions in engineering systems cannot be emphasized enough \citep{bearman1984vortex,naudascher2005flow}.

Previous studies have investigated the synchronization between the vortex shedding frequency from a bluff body, mainly for a circular cylinder, and its oscillatory motion. For various types of oscillation motions and Reynolds numbers, synchronization regimes and vortex patterns have been identified and classified in both experimental and numerical studies \citep{olinger1988nonlinear,woo1999note,ponta2005vortex,perdikaris2009chaos,leontini2011numerical,williamson1988vortex,ongoren1988flow,jacono2010modification}. These studies are based on time-consuming sweeps of the parametric space in the amplitude-frequency domain using forced motion of a bluff body. Moreover, the synchronization boundaries are identified differently depending on the identification criteria used \citep{kumar2016lock} and the wake measurement location \citep{kumar2013flow}. These subtle difficulties in the identification of precise synchronization conditions, as well as a lack of understanding the underlying physics, highlight some weaknesses of previous experimental and numerical approaches. Therefore, a theoretical approach to investigate synchronization is desirable.

The theoretical approach we consider is phase-reduction analysis, which describes the dynamics of high-dimensional periodic flows in terms of a single scalar variable ``phase." Phase-reduction analysis has been applied to nonlinear oscillators in a wide range of studies, including those on biological rhythms \citep{winfree1967biological,shiogai2010nonlinear} and chemical oscillators \citep{kuramoto2003chemical,pietras2019network}. This technique enables us to examine synchronization without extensive parametric sweeps by identifying the phase-response of flows to external perturbations \citep{taira_nakao_2018,khodkar_taira_2020,khodkar2021phase,loe2021phase}. Furthermore, phase-reduction analysis has been used to guide flow control of periodic flows \citep{nair2021phase,godavarthi_kawamura_taira_2023,loe2023controlling,fukami_nakao_taira_2024}. However, all previous applications of phase-reduction analysis were focused on two-dimensional periodic flows, even though synchronization in real engineering systems often involves three-dimensional flows. The three-dimensionality of a flow creates added richness compared to its two-dimensional counterpart, which influences the synchronization properties. Hence, it is important to identify the influence of three-dimensionality on synchronization. However, the implementation of phase-reduction analysis is not straightforward for three-dimensional wakes since, generally, they are not perfectly periodic.

In this study, we examine the effect of three-dimensionality on the synchronization of a cylinder's wake to the motion \revJK{of the cylinder}. We consider rotational, crossflow translational and streamwise translational oscillations, which are the bases of in-plane cylinder oscillation. We extend the phase reduction analysis to characterize the perturbation dynamics of three-dimensional flows. Since the underlying three-dimensionality adds fluctuations to the limit cycle, we leverage an ensemble-averaging technique to obtain mean responses of the three-dimensional wake.

This paper is organized as follows. Physical characteristics of three-dimensional wakes are presented in \S\ref{sec:FlowCharacteristics}. Fundamentals of phase-reduction analysis for periodic flows are introduced in \S\ref{sec:PhaseReductionAnalysis} with a new technique to measure the phase-response of three-dimensional flows. The results of the present phase-reduction analysis are provided in \S\ref{sec:Results}, revealing the influence of three-dimensionality \revJK{within} the flow on the synchronization. Conclusions are offered in \S\ref{sec:Conclusions}.

\section{Numerical simulation of cylinder wakes}    \label{sec:FlowCharacteristics}

\subsection{Problem description and numerical setups}

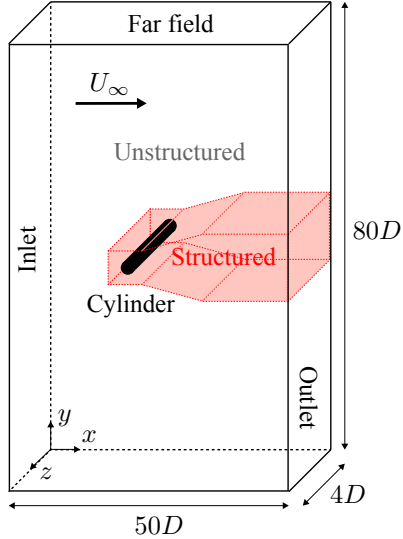
\begin{figure}
   \centering
     \includegraphics[width=0.3\textwidth]{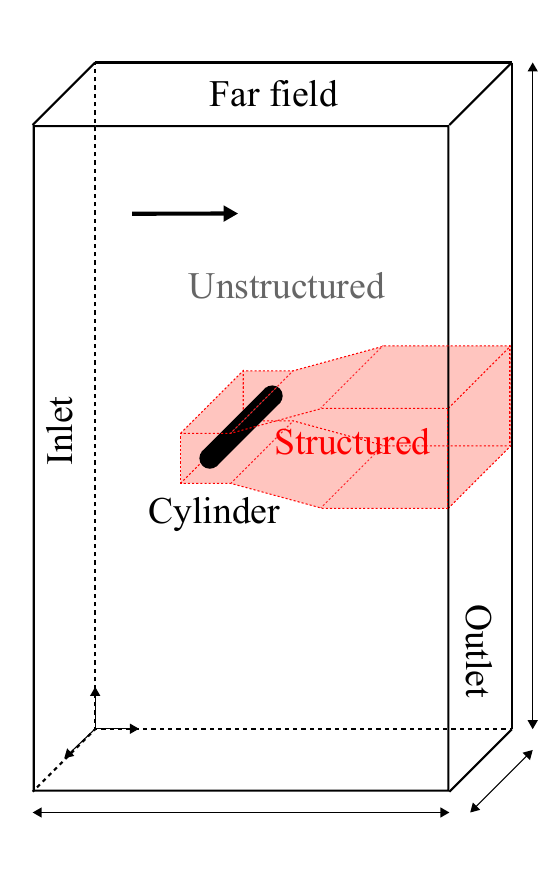}
     \put(-105,40){$x$}    \put(-114,48){$y$}    \put(-121,26){$z$}
     \put(-85,5){$50D$}   \put(-11,17){$4D$}    \put(-2.5,117){$80D$}
     \put(-102,173){$U_{\infty}$}
     \caption{Computational domain for DNS of three-dimensional cylinder wakes.} 
     \label{fig:FinalComputationalDomain}
\end{figure}

We consider two- and three-dimensional incompressible wakes past a \revJK{circular} cylinder using direct numerical simulations (DNS). The Reynolds number in this study is selected as $\Rey = U_{\infty}D/\nu = 300$, where $U_{\infty}$ is the free-stream velocity, $D$ is the diameter of the cylinder and $\nu$ is the kinematic viscosity. \revYJ{The Reynolds number is selected such that the cylinder wake develops three-dimensional structures, which is discussed in \S\ref{subsec:WakeCharacteristics} in more detail.} A finite volume formulation \citep{ham2004energy,ham2006accurate} and a fractional-step method \citep{kim1985application} with second-order accuracy are adopted for spatial discretization and time stepping, respectively. The computational domain is extended to $(x,y)/D \in [-20,30] \times [-40,40]$ from the center of the cylinder shown in figure \ref{fig:FinalComputationalDomain}. To simulate the three-dimensional wake, the spanwise extension of the computational domain is set to $z/D \in [-2,2]$ with the spanwise extent of the cylinder being $L_{z}/D = 4$, which allows us to capture the three-dimensional features related to mode A and mode B instabilities \citep{barkley1996three}. Structured grids are generated near the cylinder and in the downstream region with the uniform grid size $\Delta z/D = 0.05$ in the spanwise direction, which sufficiently resolves flow structures of mode A and B ($\lambda_{A}/D \approx 4$ and $\lambda_{B}/D \approx 0.8$). To reduce computational costs, hybrid-typed grids with approximately 0.1 million and 7.3 million volume cells are used for two- and three-dimensional wakes, respectively. Grid distributions in other directions are stretched to cluster cells near the cylinder surface. Far-field regions are discretized in an unstructured manner. The time step size of $U_\infty\Delta t /D = 0.005$ guarantees the CFL number \revJK{remains small}, $U_\infty \Delta t/\Delta x < 1$, during computations.

For the DNS of the flow over a stationary cylinder, Dirichlet boundary conditions are given to the inlet boundary with the free-stream velocity $\bs{u}=(U_{\infty},0,0)$ and the cylinder surface with $\bs{u}=\bs{0}$, respectively. The far-field boundary is prescribed with a Neumann condition of $\partial \bs{u}/\partial n = \bs{0}$. The periodic boundary condition is enforced in the spanwise direction, and a convective outlet condition is given to the outflow boundary. Our numerical solutions are validated through a comparison of the Strouhal number, $\Str$, the lift coefficient, $C_{L}$, and the drag coefficient, $C_{D}$, respectively, defined as
\begin{equation}
    \Str = \frac{f_{L}D}{U_{\infty}}, \quad C_L = \frac{F_{L}}{\frac{1}{2}\rho U_{\infty}^{2}DL_{z}}, \quad C_D = \frac{F_{D}}{\frac{1}{2}\rho U_{\infty}^{2}DL_{z}},
\end{equation}
where $\rho$ denotes the fluid density, $F_L$ and $F_D$ \revJK{represent} the lift and drag acting on the cylinder, and $f_L$ is the frequency of the lift fluctuations \revJK{that contains} the maximum energy. Table~\ref{tab:Validation_Re300} summarizes the validation of the current simulation.

\begin{table}
  \begin{center}
  \def~{\hphantom{0}}
  \begin{tabular}{lccc}

       \toprule
		                              & $\Str$ & $C^{\prime}_{L}$   & $\bar{C}_D$  \\[3pt]
       \midrule
       Present (2D)                   & 0.212    & 0.651              & 1.371        \\
       \citet{zhang1995transition}    & 0.216    & 0.692              & 1.456        \\
       \citet{persillon1998physical}  & 0.209    & 0.526              & 1.405        \\
       \citet{jiang2016three}         & 0.211    & 0.641              & 1.377        \\
                                                                                     \\
       Present (3D)                   & 0.202    & 0.42\revYJ{8}              & 1.25\revYJ{9}        \\
       \citet{zhang1995transition}    & 0.212    & 0.430              & 1.260        \\
       \citet{persillon1998physical}  & 0.206    & 0.477              & 1.366        \\
       \citet{kravchenko1999b}        & 0.203    & 0.400              & 1.280        \\
       \citet{posdziech2001numerical} & 0.201    & 0.410              & 1.250        \\
       \citet{jiang2016three}         & 0.204    & 0.464              & 1.296        \\
   	   \bottomrule
  \end{tabular}

  \hspace{300mm}

  \captionsetup{format=plain,justification=justified}
  \caption{Comparison of the Strouhal number ($\Str$), root-mean-squared lift coefficient ($C^{\prime}_{L}$), and time-averaged drag coefficients ($\bar{C}_D$) \revYJ{based on 200 shedding cycles} for two- and three-dimensional cylinder flows at $\Rey = 300$ .}
  \label{tab:Validation_Re300}
  \end{center}
\end{table}

To measure the phase-response and identify the synchronization boundaries of the cylinder wake, we introduce impulsive and oscillatory motion for the cylinder in the DNS. The motions of the cylinder are prescribed with the instantaneous translational or rotational speed of the cylinder, $U(t)$. The impulsive motion is approximated with a Gaussian function
\begin{equation}
U(t) = \varepsilon\delta(t-t_{0}) \simeq \frac{\varepsilon}{\sqrt{2\pi}\sigma}\exp\left[-\frac{1}{2}\left(\frac{t-t_0}{\sigma}\right)^2\right],
\end{equation}
where $\varepsilon$ is the magnitude of the impulsive motion and $\sigma$ determines the width of the Gaussian function, which is set \revJK{to} $\sigma=10\Delta t$ in this study. Similarly, the oscillatory cylinder motion is modeled \revJK{using the}  sinusoidal form
\begin{equation}
U(t) = U_{f} \sin(\Omega_{f}t),
\end{equation}
where $U_{f}$ and $\Omega_{f}$ are the oscillation amplitude and \revJK{angular} frequency. Cylinder motions are incorporated in the DNS by modifying the boundary conditions depending on the type of motion \citep{khodkar2021phase}. For the rotating cylinder, $U(t)$ is assigned to the tangential velocity at the cylinder surface, replacing the no-slip condition. Crossflow and streamwise translations are realized by moving the reference frame, subtracting $U(t)$ from the velocity at the inlet and far-field boundary.

\subsection{Wake characteristics of a stationary cylinder}    \label{subsec:WakeCharacteristics}

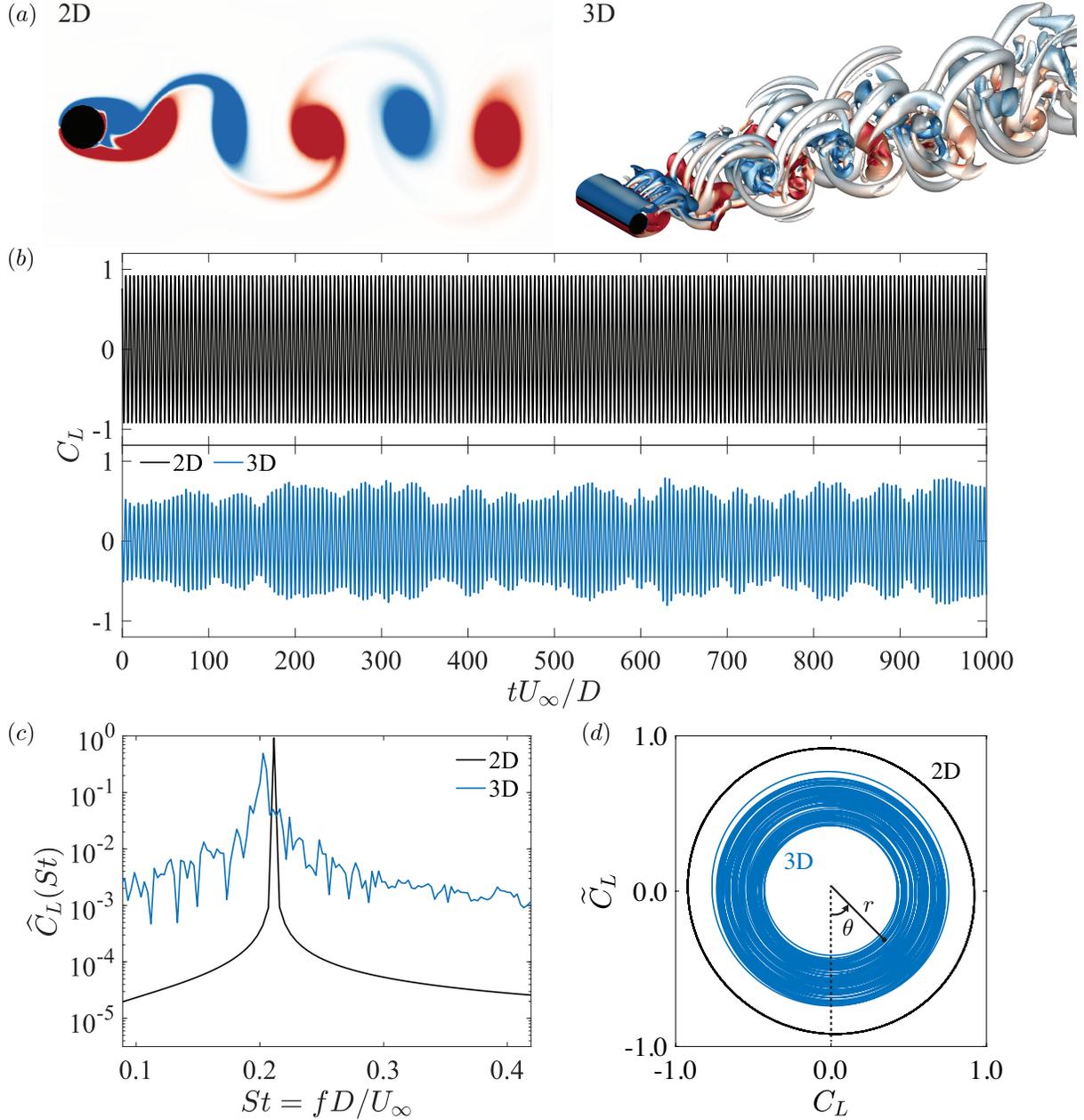
\begin{figure}
    \centering
    \includegraphics[width=\textwidth]{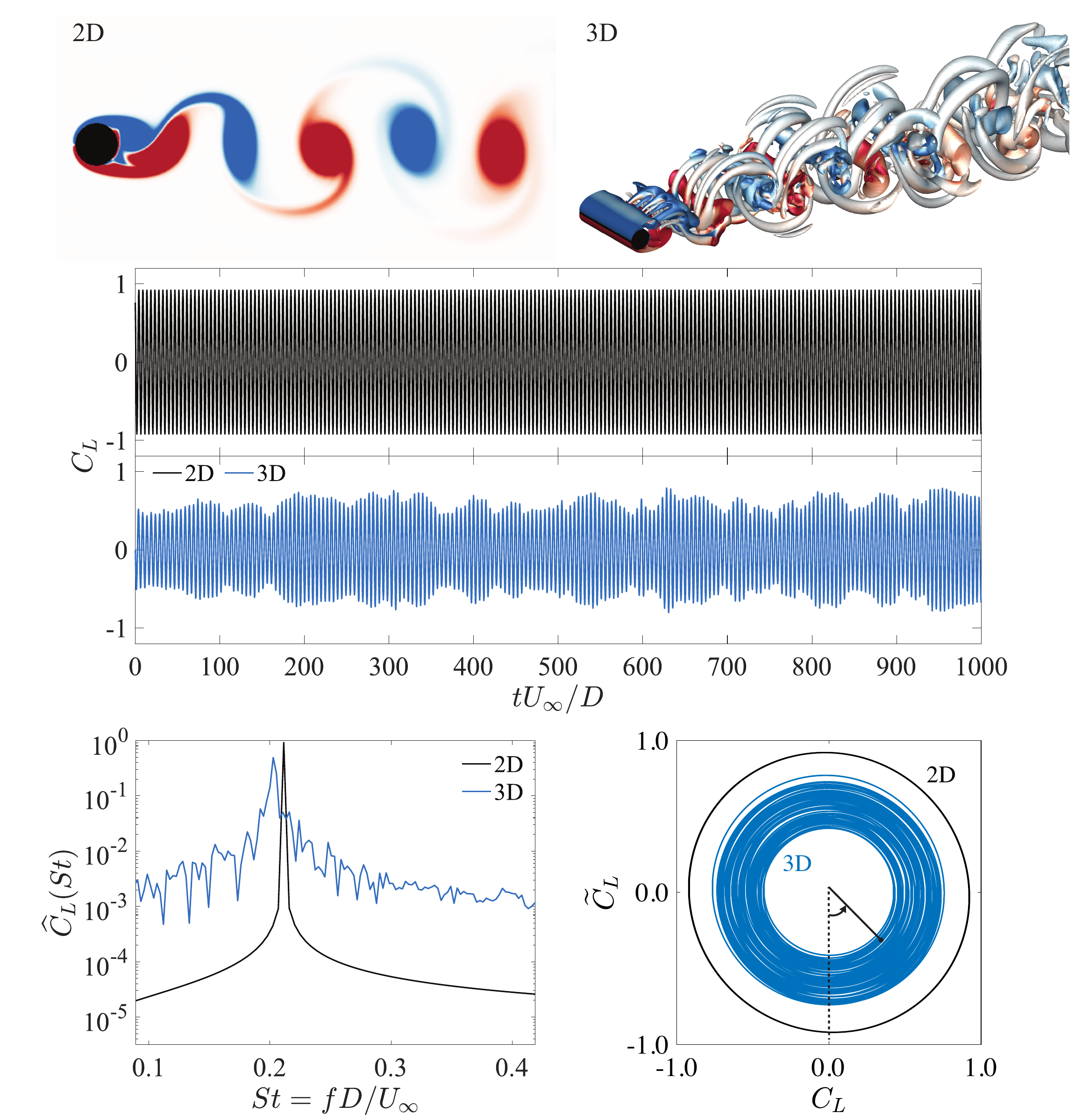}
    \put(-460,469){$(a)$}    \put(-460,365){$(b)$}    \put(-460,163){$(c)$}    \put(-216,163){$(d)$}
    \put(-96,90){$r$}    \put(-105,80){$\theta$}
    \captionsetup{format=plain,justification=justified}
    \caption{Comparison of two- and three-dimensional wakes at $\Rey$ = 300. $(a)$ Contours of spanwise vorticity of the two-dimensional wake and flow structures of three-dimensional wake visualized by the isosurface of the $Q$--criterion at $Q=0.05$ colored by spanwise vorticity. $(b)$ \revYJ{Temporal variation of the lift coefficient, $(c)$ power spectra,} $\widehat{C}_{L}$, of the lift coefficient and \revYJ{$(d)$} phase plane representation of the lift coefficient.}
    \label{fig:Comparison_2D3D}
\end{figure}

We analyze the fundamental flow physics of the three-dimensional wake of \revYJ{a stationary cylinder} at $\Rey = 300$ focusing on its representation in the phase plane. We start by providing a definition of the phase, $\theta$, and the amplitude, $r$, of a cylinder's wake based on the lift coefficient $C_{L}$ and its Hilbert transform $\widetilde{C}_{L}$ \citep{rosenblum1996phase,pikovsky1997phase} as 
\begin{equation}    \label{HilbertPhaseFormulationUCLA}
\theta(t) = \theta_{0} + \tan^{-1}\frac{\widetilde{C}_{L}(t)}{C_L(t)},\quad r\revYJ{(t)} = \sqrt{{\widetilde{C}_{L}(t)}^{2} + {C_L(t)}^{2}},
\end{equation}
\begin{equation}    \label{HilbertTransform}
\revYJ{\widetilde{C}_{L}(t) = \frac{1}{\pi} \textrm{p.v.}\int_{-\infty}^{\infty} \frac{C_L(t)}{t-\tau} \, \mathrm{d}\tau,}
\end{equation}
where \revYJ{$\textrm{p.v.}$ denotes the Cauchy principal value, and }$\theta_{0}$ is determined as $\pi/2$ to be consistent with earlier studies \citep{taira_nakao_2018,khodkar_taira_2020}.

The wake behind a circular cylinder undergoes the first bifurcation at $\Rey \approx 47$ caused by the primary instability. It triggers the wake transition from a flow with two steady counter-rotating vortices to a two-dimensional periodic flow with von K\'arm\'an vortex shedding \citep{jackson1987finite,provansal1987benard}. \revJK{Although at $\Rey = 300$, since the computational domain is restricted to only be two-dimensional for the left side graphic of figure~\ref{fig:Comparison_2D3D}$(a)$, it qualitatively shows this primary instability}. The two-dimensional wake presents a strong single peak in the \revJK{corresponding} frequency spectrum shown in figure~\ref{fig:Comparison_2D3D}\revYJ{$(c)$}, indicating the perfect periodicity of the two-dimensional wake. In the phase plane, the two-dimensional wake is represented by the limit-cycle as shown in figure~\ref{fig:Comparison_2D3D}\revYJ{$(d)$}. By increasing the Reynolds number further, \revJK{or in our case by making the computational domain three-dimensional in the case of the right side graphic of figure~\ref{fig:Comparison_2D3D}$(a)$,} the second bifurcation occurs and the wake becomes three-dimensional. This transition to the three-dimensional flow is due to the emergence of two distinct instabilities \citep{barkley1996three,williamson1988vortex,williamson1996three,posdziech2001numerical, rolandi2023compressibility}. The mode A instability occurs at $\Rey \approx 190$ with the spanwise wavelength $\lambda_{A}/D \approx 4$ which deforms the primary vortices. The mode B instability starts to appear at the higher Reynolds number of $\Rey\approx 240$ with the spanwise wavelength $\lambda_{B}/D \approx 0.8$. This instability mode is associated with secondary fine structures in the braid region, stretched in the streamwise direction and connecting the primary vortices. These instability modes create \revYJ{time-varying lift fluctuations associated with the }more broadband frequency spectrum \revJK{and} a corresponding shift in the peak frequency\revYJ{,} as shown in figure~\ref{fig:Comparison_2D3D}\revYJ{$(b)$} and \revYJ{$(c)$}. The three-dimensional wake exhibits an unclosed trajectory with the amplitude variation in the phase plane, depicted in figure~\ref{fig:Comparison_2D3D}\revYJ{$(d)$}.

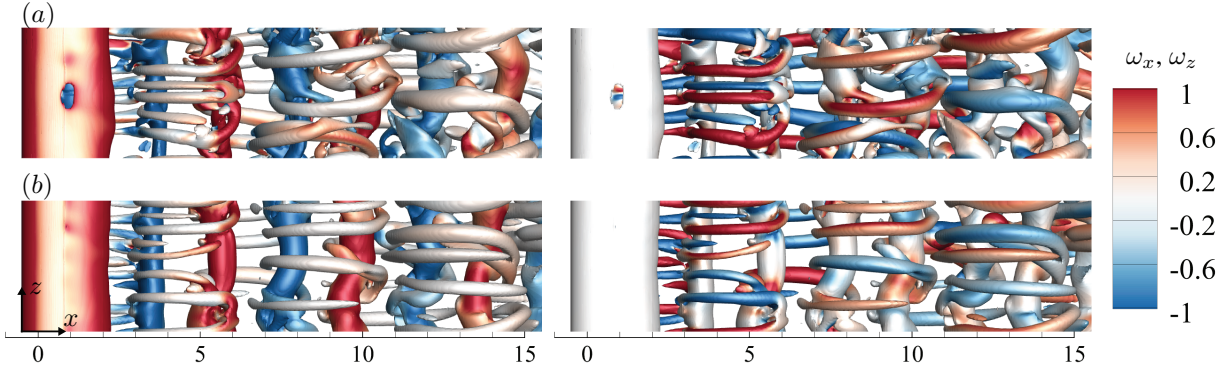
\begin{figure}
    \centering
    \includegraphics[width=\textwidth]{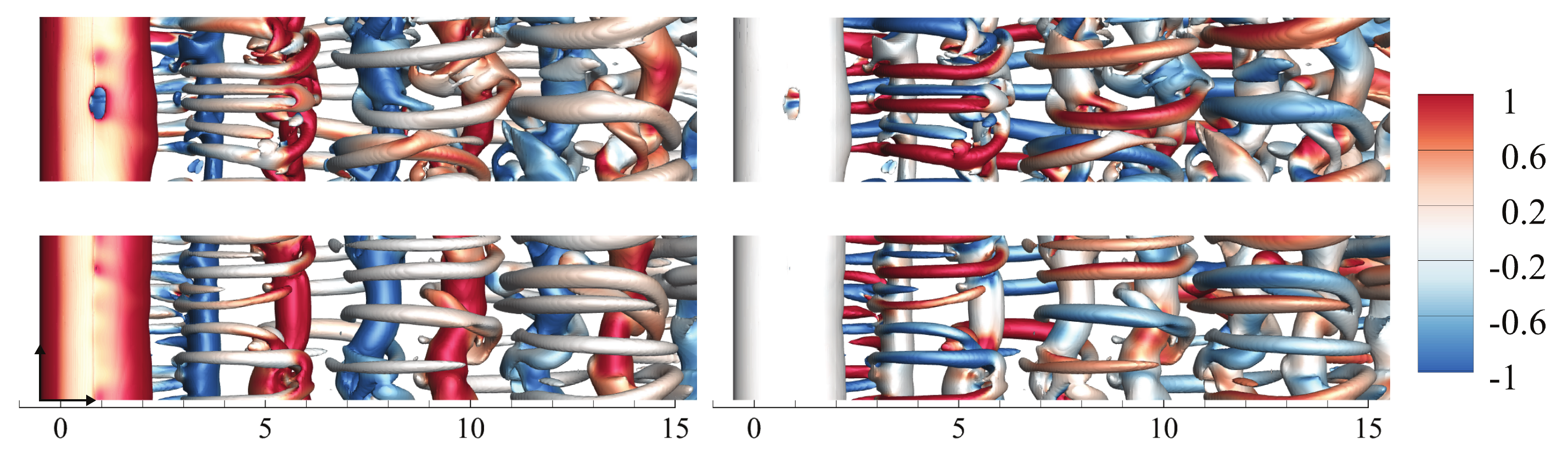}
    \put(-442,18){$x$}    \put(-455,30){$z$} 
    \put(-40,117){$\omega_{x}$, $\omega_{z}$}
    \put(-458,133){$(a)$}    \put(-458,68){$(b)$}
    \captionsetup{format=plain,justification=justified}
    \caption{Instantaneous \revYJ{wake from a stationary cylinder }visualized by isosurfaces of the $Q$-criterion ($Q=0.05$) colored with $\omega_z$ (left column) and $\omega_x$ (right column) corresponding to $(a)$ $r=0.44$ and $(b)$ $0.73$.}
    \label{fig:FlowStructure}
\end{figure}

Two instantaneous flow snapshots corresponding to lower and higher amplitudes, $r=0.44$ and $0.73$, at the same phase, are visualized in figures~\ref{fig:FlowStructure}$(a)$ and $(b)$, respectively. The visualizations distinctly emphasize the primary and secondary vortical structures by coloring the spanwise vorticity, $\omega_{z}$, and streamwise vorticity, $\omega_{x}$. We observe that the secondary vortical structures are more uniformly arranged and aligned with each other in the higher amplitude cycle with $r = 0.73$. Moreover, the primary vortices are more coherent along the spanwise direction. Contrarily, in the lower amplitude cycle with $r = 0.44$, the streamwise stretched vortical structures appear less organized, and the primary vortex cores exhibit greater dislocations and deformation along the spanwise direction.

\begin{figure}
    \centering
    \includegraphics[width=\textwidth]{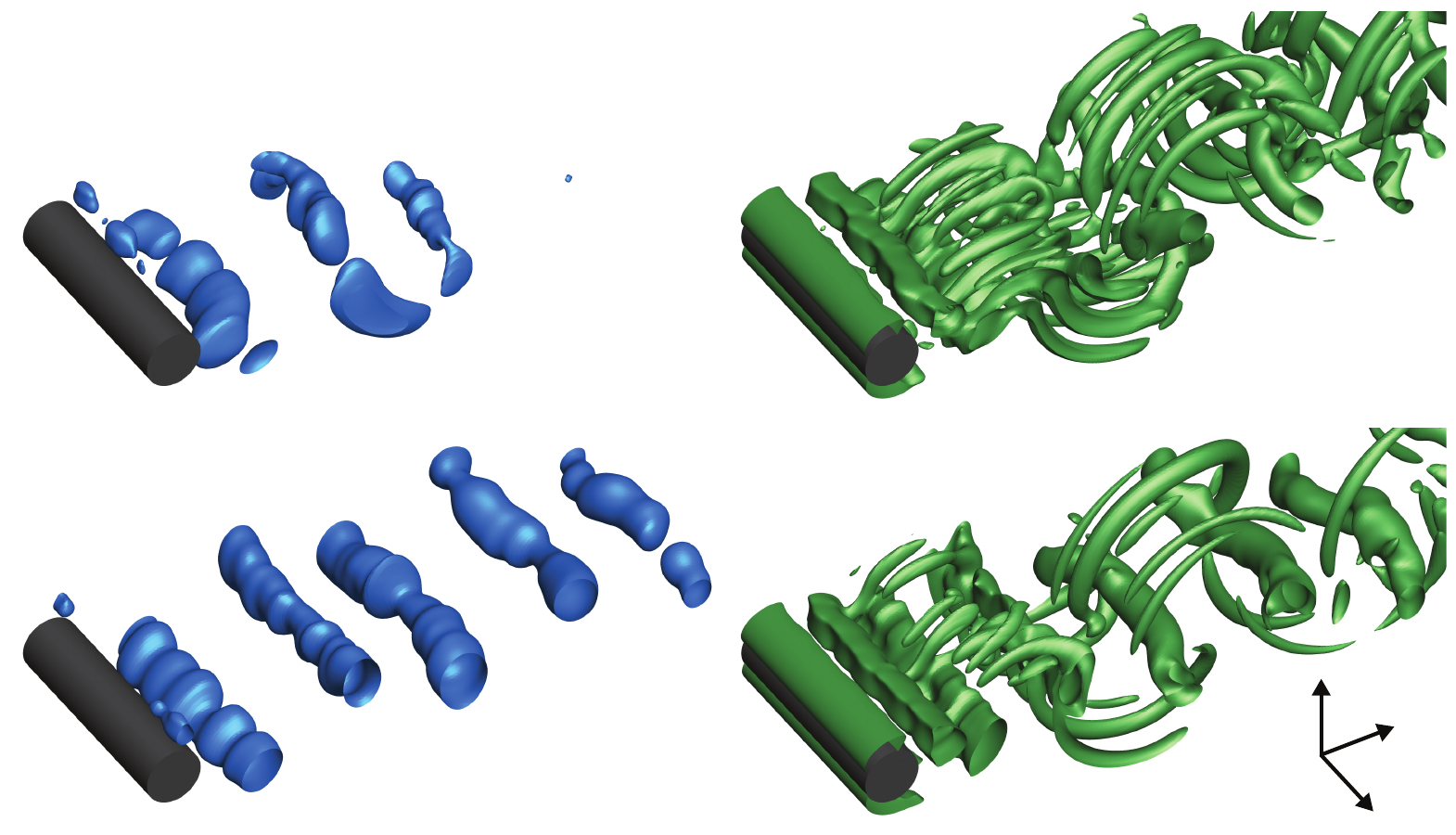}
    \put(-21,25){$x$}    \put(-38.5,46){$y$}    \put(-24,8){$z$} 
    \put(-463,258){$(a)$}    \put(-463,123){$(b)$}
    \captionsetup{format=plain,justification=justified}
    \caption{Instantaneous visualization of scale-decomposed flows by isosurfaces of the $Q$-criterion ($Q=0.05$) corresponding to $(a)$ $r=$ 0.44 and $(b)$ 0.73. The left and right columns \revJK{capture} large-scale $[\sigma_{\max},2\sigma_{\max}]$ and small-scale $[\sigma_{\max}/4,\sigma_{\max}/2]$ structures, respectively.}
    \label{fig:FlowStructure_ScaleDecomposed}
\end{figure}

Similar trends are seen more clearly in scale-decomposed flow structures \citep{fujino2023hierarchy}. We apply a spatial band-pass filter to the velocity field based on a two-dimensional Gaussian kernel, $G$, in the $x$ and $y$ directions with the length scale range of $[\sigma_{1},\sigma_{2}]$ as
\begin{equation}
\bs{u}^{[\sigma_{1},\sigma_{2}]}(x,y,z,t) = \int_{\mathcal{S}} \bs{u}(x,y,z,t)\left[G(x',y';x,y,\sigma_{1})-G(x',y';x,y,\sigma_{2})\right] \, \mathrm{d}x'\mathrm{d}y',
\end{equation}
where
\begin{equation}
G(x',y';x,y,\sigma) = \frac{1}{2\pi\sigma^{2}}\exp\left[-\frac{(x-x')^2 + (y-y')^2}{2\sigma^2}\right]
\end{equation}
and the integration window $\mathcal{S}$ covers the whole computational domain in the $x$ and $y$ directions. Large- and small-scale flow structures with a length scale range of $[\sigma_{\max},2\sigma_{\max}]$ and $[\sigma_{\max}/4,\sigma_{\max}/2]$ are visualized in figure~\ref{fig:FlowStructure_ScaleDecomposed}, where the cutoff length scale is $\sigma_{\max}=D/(2\pi \Str)$ \citep{yasuda2020formation}. We observe more large-scale structures as the amplitude, $r$, increases, while the small-scale structures diminish. These observations tell us that the cylinder wake is more irregular as the amplitude, $r$, decreases.

\begin{figure}
    \centering
    \includegraphics[width=\textwidth]{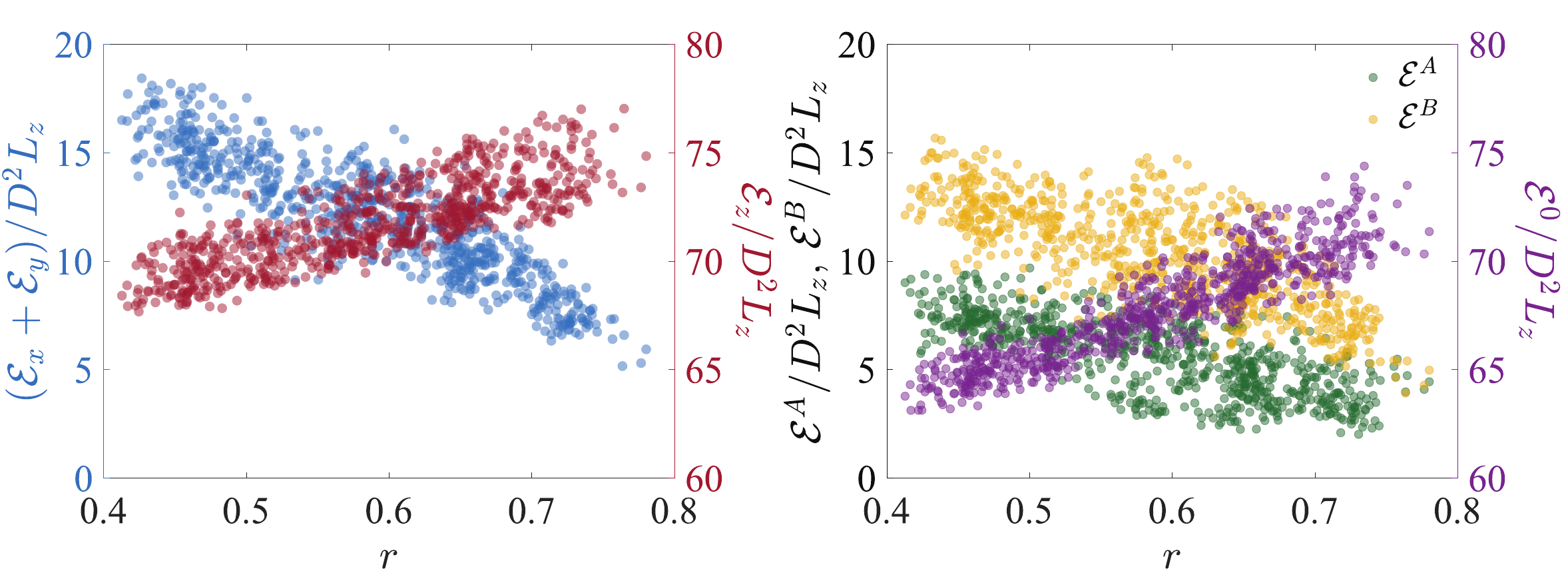}
    \put(-470,158){$(a)$}    \put(-235,158){$(b)$}
    \captionsetup{format=plain,justification=justified}
    \caption{Variation of enstrophies with $r$. (a) Spanwise enstrophy $\mathcal{E}_z$ and non-spanwise enstrophy $\mathcal{E}_x+\mathcal{E}_y$. (b) Enstrophies of spanwise-averaged vorticity $\mathcal{E}^{0}$, mode A $\mathcal{E}^{A}$ and mode B $\mathcal{E}^{B}$.}
    \label{fig:Enstrophy}
\end{figure}

These patterns can also be quantitatively identified by enstrophies defined as
\begin{equation}
\mathcal{E}_{i} = \int_{\mathcal{V}} \omega_{i}^{2} \, \mathrm{d}\bs{x},\quad\mathcal{E} = \int_{\mathcal{V}} \bs{\omega} \cdot \bs{\omega} \, \mathrm{d}\bs{x} = \mathcal{E}_{x} + \mathcal{E}_{y} + \mathcal{E}_{z},
\end{equation}
where $\omega_{i}$ denotes the vorticity component in each direction ($i=x$, $y$, and $z$). The non-spanwise enstrophy $\mathcal{E}_{x}+\mathcal{E}_{y}$ characterizes the three-dimensionality of the wake induced by the spanwise mixing. The integration window $\mathcal{V}$ is set to $(x,y,z)/D \in [-1,4] \times [-3,3]\times [-2,2]$ to examine flow structures formed within one shedding period. The variation in the non-spanwise enstrophy, $\mathcal{E}_x+\mathcal{E}_y$, and spanwise enstrophy, $\mathcal{E}_z$, with $r$ is shown in figure~\ref{fig:Enstrophy}$(a)$. A decrease in $r$ corresponds to a decrease in $\mathcal{E}_z$ and an increase in $\mathcal{E}_x+\mathcal{E}_y$. This indicates that larger spanwise mixing occurs when $r$ is smaller and induces enhancement in three-dimensional flow structures. Moreover, weaker spanwise vortices are reflected \revJK{as portions of smaller fluctuation amplitude in the $C_L$ time history}, which corresponds to smaller $r$ \revJK{in the phase plane}. 

\begin{figure}
    \centering
    \includegraphics[width=\textwidth]{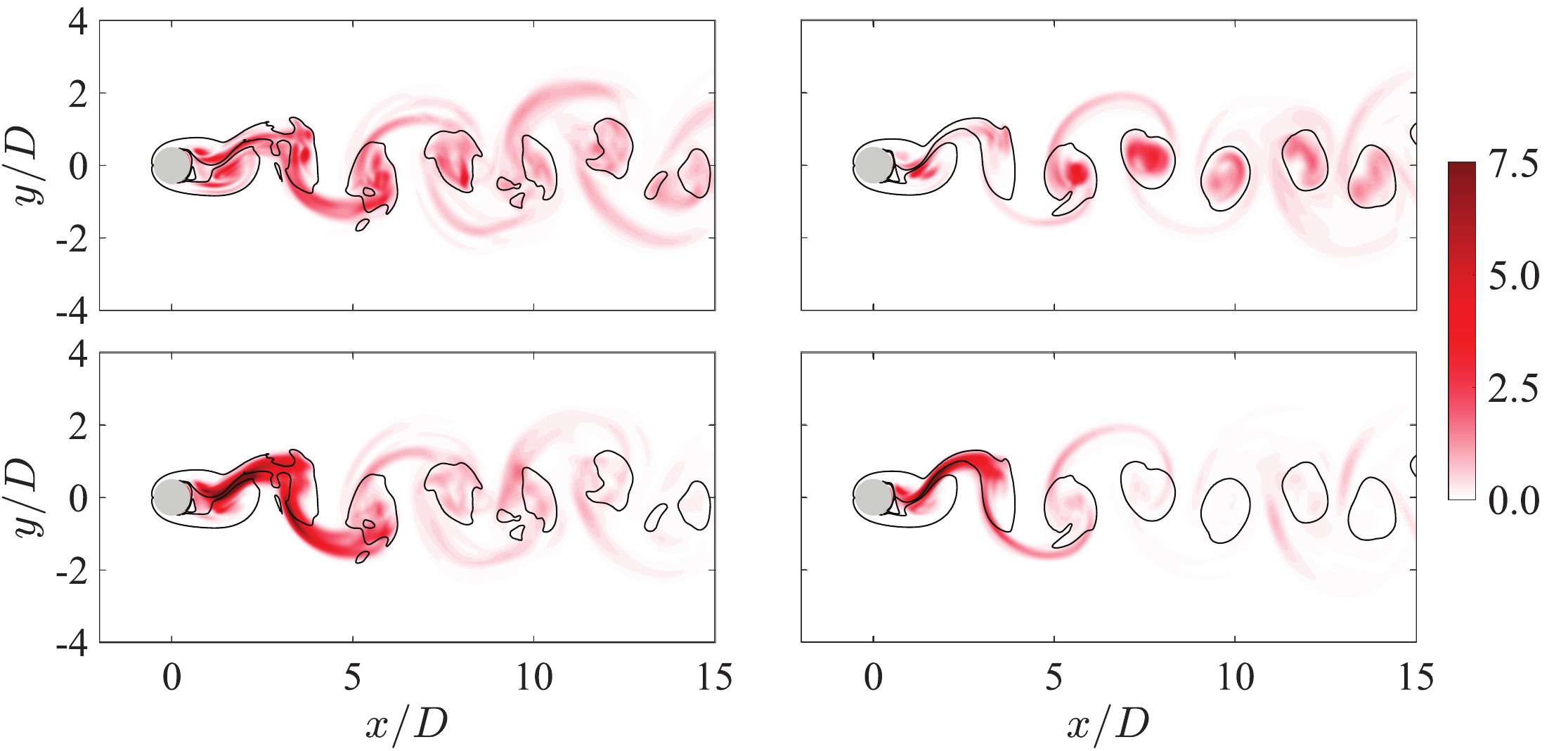}
    \put(-469,217){$(a)$}    \put(-248,217){$(b)$}
    \captionsetup{format=plain,justification=justified}
    \caption{Contours of spanwise averaged $\bs{\omega}^{A} \cdot \bs{\omega}^{A}$ (top row) and $\bs{\omega}^{B} \cdot \bs{\omega}^{B}$ (bottom row) for $(a)$ $r=0.44$ and $(b)$ $0.73$. The black lines represent the contour lines of $\omega^{0}_{z} = \pm 0.5$.}
    \label{fig:FlowStructure_Enstrophy}
\end{figure}

To identify the dependencies between the instabilities and phase-amplitude variations of the three-dimensional wake, we apply the discrete Fourier transform (DFT) in the spanwise direction, which leads to the decomposition of the vorticity field as
\begin{equation}    \label{FFT}
\bd{\omega}(\bs{x},t) = \sum_{m=0}^{N_z/2} \widehat{\bs{\omega}}(x,y,t,m)e^{2\pi imz/L_{z} } = \bs{\omega}^{0}(\bs{x},t) + \bs{\omega}^{A}(\bs{x},t) + \bs{\omega}^{B}(\bs{x},t), 
\end{equation}
where $\widehat{\bs{\omega}}$ denotes the coefficient of each Fourier mode and $N_z = L_z / \Delta z$ is the number of grid points in the spanwise direction. We categorize the Fourier modes into three vorticity components corresponding to the spanwise average ($\bs{\omega}^{0}$), mode A ($\bs{\omega}^{A}$) and mode B ($\bs{\omega}^{B}$). In other words, Fourier modes with large wavelengths, such that $\lambda/D \geq 2$, are considered mode A, and the remaining  Fourier modes, with shorter wavelengths, are classified as mode B. This wavelength threshold is chosen \revJK{to fall} within a range of stable wavelengths that separate the range of unstable wavelengths of mode A and mode B \citep{leontini2015stability,rolandi2023compressibility}.

As in the previous analysis, we quantify the contribution of each vorticity component by evaluating total enstrophy as shown in figure~\ref{fig:Enstrophy}$(b)$. Attributing to the orthogonality of Fourier modes, the total enstrophy, $\mathcal{E}$, is also decomposed into three parts as $\mathcal{E} = \mathcal{E}^{0} + \mathcal{E}^{A} + \mathcal{E}^{B}$. Mode B surpasses mode A through the entire range of $r$, which implies the finer structures from the mode B instability are more prominent than large-scale structures from mode A. Both $\mathcal{E}^{A}$ and $\mathcal{E}^{B}$ become larger as $r$ decreases, whereas $\mathcal{E}^{0}$ shows the opposite trend. Thus, it can be argued that secondary instabilities convert two-dimensional flow features in the wake to three-dimensional structures as $r$ decreases.

We also identify regions where each instability mode actively grows by plotting the magnitudes of each vorticity component in figure~\ref{fig:FlowStructure_Enstrophy}. For the shedding cycle with $r = 0.73$, the mode A vorticity component highlights primary vortex cores, while mode B structures reside in the shear layers connecting two primary vortices. Flow structures of mode A last longer in the downstream region than mode B structures which implicates a faster decay rate of the mode B structures \citep{jiang2016three}. Moreover, smooth isocountour lines of $\omega_{z}^{0}$ in figure~\ref{fig:FlowStructure}$(b)$ reflect the coherence of primary vortices in the spanwise direction. These trends are generally maintained in shedding cycles with smaller $r$, although the distributions of instability modes become more irregular due to the enhanced spanwise mixing. In particular, we observe that the mode A vorticity component at smaller $r$ no longer only highlights primary vortex core regions but it is also present in the braid region. Flow structures associated with the mode B vorticity component are stronger and thickened in the braid regions, which gives rise to larger elongated vortical structures in this region. This can also be seen in figure \ref{fig:FlowStructure}. These originate from the thickening of the shear layers in the braid region connecting the two primary vortices, which now allows flow structures with the larger wavelength to be amplified within this region.

\section{Phase-reduction analysis of fluid flows}    \label{sec:PhaseReductionAnalysis}

\subsection{Phase-reduction analysis of periodic flows}
\label{sec:PhaseReductionAnalysis_Periodic}

The dynamics of incompressible flows governed by the Navier--Stokes (NS) equations can be expressed with the state vector $\bs{q}$ and the NS operator $\bs{\mathcal{N}}$ as 
\begin{equation}    \label{eq:GoverningEq}
    \frac{\partial}{\partial t} \bs{q}(\bs{x},t) = \bs{\mathcal{N}}(\bs{q}(\bs{x},t)),
\end{equation}
where $\bs{x}$ and $t$ denote dimensionless space and time, respectively. A periodic flow can be described as a limit cycle solution $\bs{q}_0$ of the governing equation \ref{eq:GoverningEq} with phase $\theta$. Since $\bs{q}_0$ is periodic, $\bs{q}_0(\bs{x},t) = \bs{q}_0(\bs{x},t+2\pi/\Omega_{n})$ where $\Omega_n$ is the natural \revYJ{angular} frequency of the flow. Here, we consider the phase functional $\Theta(\bs{q})$ that allows us to relate the high-dimensional instantaneous flow field $\bs{q}$ to its phase variable $\theta$. The concept of phase, $\Theta(\bs{q})$, is then extended to the vicinity of the limit cycle to describe weakly perturbed flows. More specifically, if the flow $\bs{q}(\bs{x},t)$ asymptotically converges to the limit cycle solution $\bs{q}_{0}(\bs{x},t)$ as it develops, the phase of $\bs{q}$ is defined as $\Theta(\bs{q}(\bs{x},t)) = \Theta(\bs{q}_{0}(\bs{x},t))$.
Combined with the governing equation \ref{eq:GoverningEq}, the phase dynamics of the unperturbed periodic flow can be described using functional derivatives of $\Theta(\bs{q})$ as
\begin{equation}
    \dot{\theta} = \dot{\Theta}(\bs{q}(\bs{x},t))
                 = \int_{\mathcal{V}_f} \frac{\delta\Theta}{\delta\bs{q}} \cdot \dot{\bs{q}} \, \mathrm{d\bs{x}}
                 = \int_{\mathcal{V}_f} \frac{\delta\Theta}{\delta\bs{q}} \cdot \bs{\mathcal{N}}(\bs{q}) \, \mathrm{d}\bs{x}
                 = \Omega_{n},
\label{eq:PhaseDynamics_woPerturbation}
\end{equation}
where $\mathcal{V}_f$ denotes the control volume. 

Now, suppose a small perturbation term, with magnitude $\varepsilon<1$ and spatiotemporal profile $\bs{p}(\bs{x},t)$ such that $\lVert \bs{p}(\bs{x},t) \rVert = 1$, is added to equation \ref{eq:GoverningEq} to analyze the dynamics of perturbed flows,
\begin{equation}    \label{eq:GoverningEq_Perturbed}
    \frac{\partial}{\partial t} \bs{q}(\bs{x},t) = \bs{\mathcal{N}}(\bs{q}(\bs{x},t)) + \varepsilon\bs{p}(\bs{x},t).
\end{equation}
This leads to the phase dynamics of weakly perturbed flows,
\begin{equation}    \label{eq:PhaseDynamics_wPerturbation}
    \dot{\theta} = \int_{\mathcal{V}_f} \frac{\delta\Theta}{\delta\bs{q}} \cdot (\bs{\mathcal{N}}(\bs{q}) + \varepsilon\bs{p}(\bs{x},t)) \, \mathrm{d}\bs{x}
                 = \Omega_{n} + \varepsilon\int_{\mathcal{V}_f} \frac{\delta\Theta}{\delta\bs{q}} \cdot \bs{p}(\bs{x},t) \, \mathrm{d}\bs{x}.
\end{equation}
For a weak perturbation, the functional derivative $\delta\Theta/\delta\bs{q}$ can be linearly approximated with respect to the limit cycle $\bs{q}=\bs{q}_{0}$. This yields the spatial phase-sensitivity function
\begin{equation}    \label{eq:SpatialPhaseSensitivity}
\bs{Z}(\bs{x};\theta) \equiv \left. \frac{\delta\Theta(\bs{q})}{\delta\bs{q}}\right|_{ \bs{q}=\bs{q}_{0}},
\end{equation}
which provides the linear phase-response of fluid flows to small external perturbations. In particular, equation~\ref{eq:PhaseDynamics_wPerturbation} is further simplified by introducing the phase-sensitivity function $\zeta(\theta)$ associated with the spatial forcing profile $\bs{f}(\bs{x})$ as
\begin{equation}    \label{eq:PhaseSensitivityFunction}
        \dot{\theta} = \Omega_{n} + \varepsilon\eta(t)\zeta(\theta), \quad \zeta(\theta) = \int_{\mathcal{V}_f} \bs{Z}(\bs{x};\theta) \cdot \bs{f}(\bs{x}) \, \mathrm{d}\bs{x},
\end{equation}
when the perturbation can be decomposed as $\bs{p}(\bs{x},t) = \eta(t)\bs{f}(\bs{x})$.

Among the various approaches to determine the phase-sensitivity function $\zeta(\theta)$ \citep{nakao2016phase,iima2019jacobian,kawamura2022adjoint}, we employ the direct method which can be performed in both numerical and experimental studies \citep{taira_nakao_2018}. For this method, a weak impulsive perturbation $\eta(t) = \delta(t-t_{0})$ is applied to the flow at different phases, where $\delta(t-t_{0})$ is a Dirac-delta function centered at $t=t_{0}$. When the transient effects of the impulsive perturbation settle, the perturbed flow returns to a stable state on the limit cycle, leaving an asymptotic phase shift. The asymptotic phase shift is often referred to as the phase-response function $g(\theta;\varepsilon) = \lim_{t\to\infty} [\Theta(\bs{q}_{\delta}(\bs{x},t)) - \Theta(\bs{q}_{0}(\bs{x},t))]$, where $\bs{q}_{\delta}$ denotes the state vector of the flow impulsively perturbed at $\theta$. The phase-sensitivity function, $\zeta(\theta)$, is calculated by integrating equation~\ref{eq:PhaseSensitivityFunction} with an impulsive perturbation and results in $\zeta(\theta) \approx g(\theta;\varepsilon)/\varepsilon$ through the first-order approximation \citep{nakao2016phase}. \revYJ{Based on this perspective, the phase-sensitivity function can be physically interpreted as how much phase advancement, or delay, would be induced by imposing an external impulsive perturbation at a particular phase of the system.}

Using equation~\ref{eq:PhaseSensitivityFunction}, the synchronization of wake dynamics to any periodic forcing, with \revYJ{angular} frequency $\Omega_f$, can be investigated theoretically. Synchronization between the flow and the periodic forcing is achieved when the relative phase $\phi(t) \equiv \theta(t) - \Omega_{f}t/m$ converges to a constant value, i.e.,
\begin{equation}    \label{eq:PhaseDifference}
    \dot{\phi} = \Omega_{n} - \frac{\Omega_{f}}{m} + \varepsilon\eta(t)\zeta\left(\phi + \frac{\Omega_{f}}{m} t \right) \rightarrow 0,
\end{equation}
where the natural number $m$ denotes the subharmonic number of interest. This expression can be rewritten in an autonomous form by averaging over a period of the perturbation $T_{f} = 2\pi/\Omega_{f}$ \citep{kuramoto2003chemical,ermentrout2010mathematical},
\begin{equation}    \label{eq:PhaseDifferenceAutonomous}
    \dot{\phi} \simeq \Omega_{n} - \frac{\Omega_{f}}{m} + \varepsilon \mathit{\Gamma}(\phi),
\end{equation}
where $\mathit{\Gamma}(\phi)$ is the phase-coupling function given by
\begin{equation}    \label{eq:PhaseCouplingFunction}
    \mathit{\Gamma}(\phi) = \frac{1}{T_{f}} \int _{-\pi/\Omega_{f}} ^{\pi/\Omega_{f}} \zeta\left(\phi + \frac{\Omega_{f}}{m}\tau \right) \eta(\tau) \, \mathrm{d}\tau
                 = \frac{1}{2\pi } \int _{-\pi} ^{\pi}  \zeta\left(\phi + \frac{\varphi}{m}\right) \eta\left(\frac{\varphi}{\Omega_{f}}\right) \, \mathrm{d}\varphi.
\end{equation}
Considering the stable solution of equation \ref{eq:PhaseDifferenceAutonomous}, the condition for synchronization can be theoretically derived as
\begin{equation}    \label{eq:SynchronizationCondition}
    \varepsilon \min \mathit{\Gamma}(\phi) \leq \frac{\Omega_{f}}{m} - \Omega_{n} \leq \varepsilon \max \mathit{\Gamma}(\phi),
\end{equation}
and the synchronizability is defined as $S = \max \mathit{\Gamma} - \min \mathit{\Gamma}$. The condition described by equation~\ref{eq:SynchronizationCondition} theoretically identifies the possible combination map of the forcing \revYJ{angular} frequency, $\Omega_{f}$, and magnitude, $\varepsilon$, that can achieve synchronization, known as the Arnold tongue. The synchronizability, $S$, conveys the ease by which synchronization is achieved for a specific periodic perturbation at a different forcing frequency. A larger $S$ indicates that a smaller motion is sufficient for the flow to be synchronized with external forcing at \revYJ{angular} frequency $\Omega_f$.

\subsection{Ensemble-based phase-reduction framework}

Strictly speaking, the phase-reduction framework is only valid for perfectly periodic flows. However, as discussed in \S\ref{sec:FlowCharacteristics}, three-dimensional flows do not exhibit perfect limit-cyclic behavior which requires care when performing phase-reduction analysis. In recent years, a few studies have worked on extending the phase-reduction approach to chaotic oscillators \citep{josic2001phase,schwabedal2012optimal,kurebayashi2012theory,tonjes2022phase,imai2022phase}. Inspired by these works, we extend the phase-based description to three-dimensional flows.

The phase dynamics of three-dimensional flows can be expressed as
\begin{equation}
\label{eq:PhaseDynamics_woPerturbation_PseudoPeriodic}
    \dot{\theta} = \int_{\mathcal{V}_f} \frac{\delta\Theta}{\delta\bs{q}} \cdot \bs{\mathcal{N}}(\bs{q}) \, \mathrm{d}\bs{x} = \Omega_{n} + F(\bd{q}(\bd{x},t)),\quad
    \int_{-\infty}^{\infty} F(\bd{q}(\bd{x},t)) \,\mathrm{d}t = 0,
\end{equation}
where the term $F(\bd{q}(\bd{x},t))$ represents the instantaneous frequency fluctuation from the natural \revYJ{angular} frequency $\Omega_{n}$, which stems from the phase diffusion of systems \citep{josic2001phase,kurebayashi2012theory}. Following the same procedure provided in section~\ref{sec:PhaseReductionAnalysis_Periodic}, the phase dynamics of the weakly perturbed three-dimensional flows are modeled as
\begin{equation}    \label{eq:PhaseDynamics_wPerturbation_PseudoPeriodic}
    \dot{\theta} = \Omega_{n} + F(\bd{q}(\bd{x},t)) + \varepsilon\int_{\mathcal{V}_f} \frac{\delta\Theta}{\delta\bs{q}} \cdot \bs{p}(\bs{x},t) \, \mathrm{d}\bs{x}.
\end{equation}
Since we are more interested in the long-term dynamics when investigating synchronization, observing the mean phase dynamics of the flow is more valuable than the instantaneous dynamics. Hence, we apply an ensemble-averaging to equation~\ref{eq:PhaseDynamics_wPerturbation_PseudoPeriodic} which can be expressed as
\begin{equation}    \label{eq:EnsembleAverage}
    \langle X \rangle = \lim_{K\to\infty} \frac{1}{K}\sum_{k=1}^{K} X^{(k)},
\end{equation}
where $\langle\cdot\rangle$ denotes the ensemble-averaging operator and $X^{(k)}$ is the $k$-th realization of a variable $X$. This yields
\begin{equation}    \label{eq:PhaseDynamics_wPerturbation_EnsembleAverage}
    \langle\dot{\theta}\rangle = \Omega_{n} + \varepsilon \int_{\mathcal{V}_f} \left\langle \frac{\delta\Theta}{\delta\bs{q}} \right\rangle \cdot \bs{p}(\bs{x},t) \, \mathrm{d}\bs{x},
\end{equation}
which has the same form as equation \ref{eq:PhaseDynamics_wPerturbation}. Thus, we define the spatial phase-sensitivity function of three-dimensional flows as
\begin{equation}    \label{eq:PhaseSensitivity_EnsembleAveraged}
    \bs{Z}(\bs{x};\theta) =  \left\langle \left. \frac{\delta\Theta(\bs{q})}{\delta\bs{q}}  \right|_{\scriptscriptstyle \bs{q}=\bs{q}_{0}} \right\rangle,
\end{equation}
which evaluates $\langle \delta\Theta/\delta\bs{q} \rangle$ from equation \ref{eq:PhaseDynamics_wPerturbation_EnsembleAverage} with respect to the unperturbed flow $\bs{q}_{0}(\bs{x},t)$. We then obtain the phase-sensitivity function $\zeta(\theta)$ of three-dimensional flows for a given spatial forcing profile $\bs{f}(\bs{x})$ according to equation~\ref{eq:PhaseSensitivityFunction}. This leads us to the identical formulation of phase-coupling functions and synchronization conditions provided in equations~\ref{eq:PhaseCouplingFunction} and~\ref{eq:SynchronizationCondition} for three-dimensional flows.
\begin{figure}
    \centering
    \includegraphics[width=400pt]{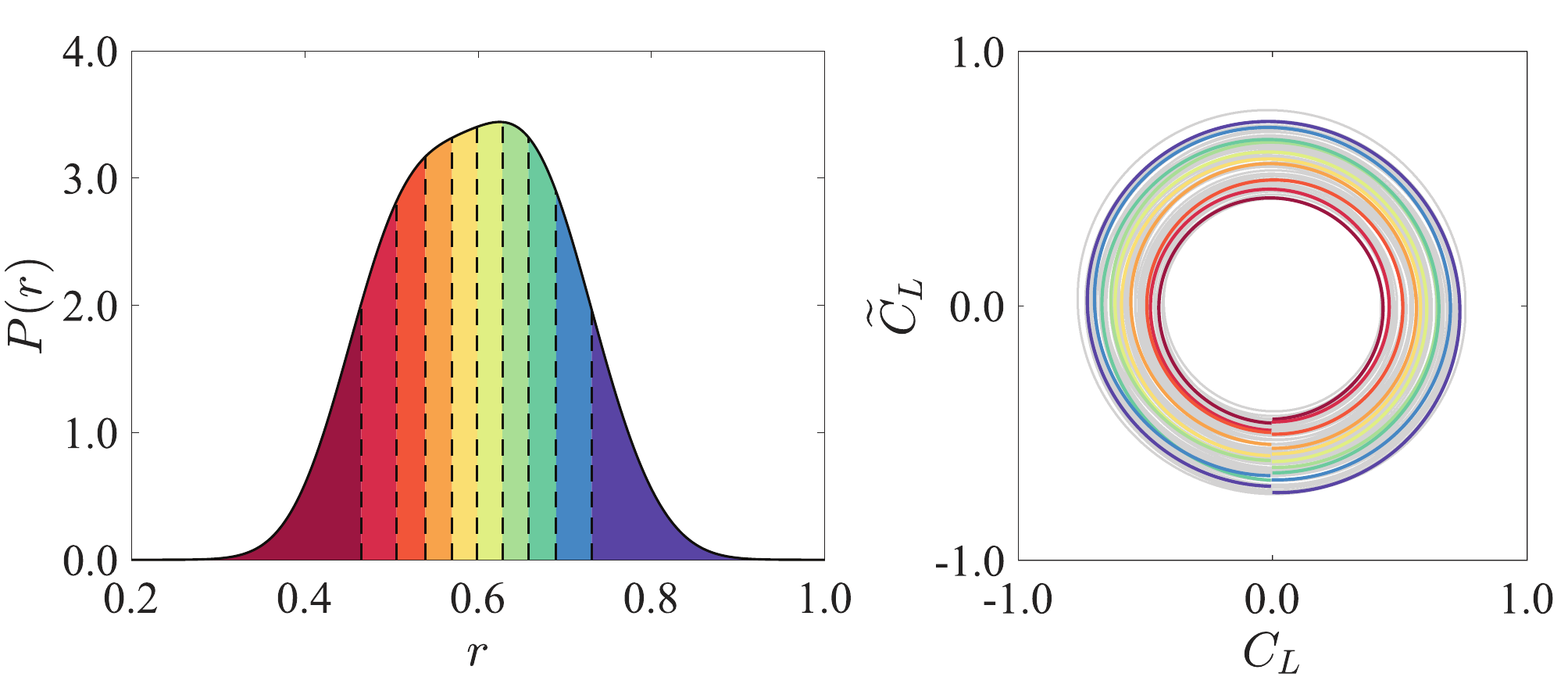}
    \put(-400,159){$(a)$}    \put(-178,159){$(b)$}
    \captionsetup{format=plain,justification=justified}
    \caption{$(a)$ Probability density function of cycle amplitude $r$ for extraction and $(b)$ 10 selected sample cycles.} 
    \label{fig:Sampling}
\end{figure}

\begin{figure}
    \centering
    \includegraphics[width=\textwidth]{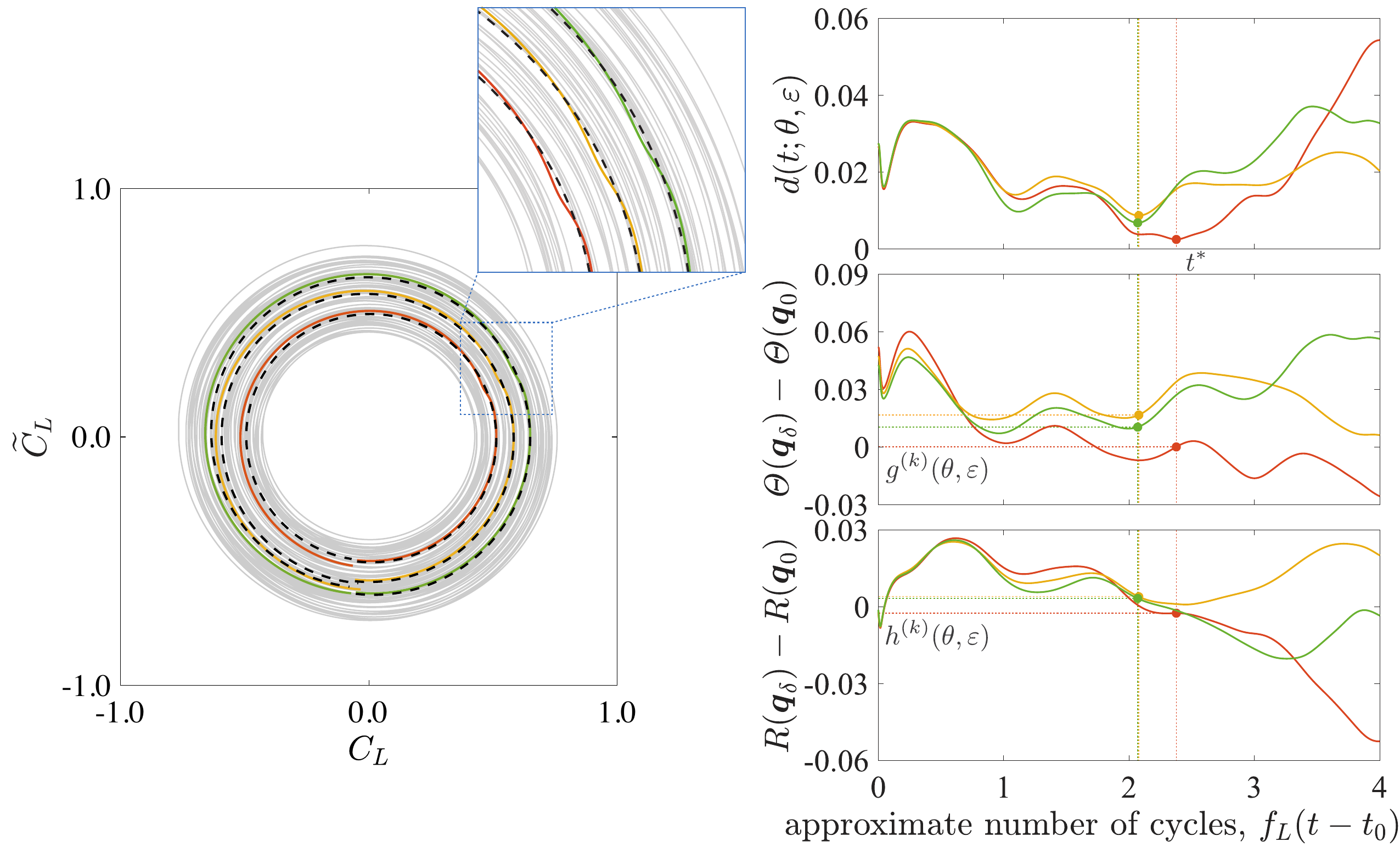}
    \put(-463,218){$(a)$}    \put(-215,275){$(b)$}
    \captionsetup{format=plain,justification=justified}
    \caption{($a$) Representation \revJK{in the phase plane} of three \revYJ{sample cycles, each with different amplitudes,} perturbed at \revYJ{$\theta=0.6\pi$ with rotary impulsive cylinder motion }(solid) and \revJK{the corresponding} unperturbed sample cycles (\revYJ{dashed}). ($b$) Estimation of $g^{(k)}(\theta,\varepsilon)$ and $h^{(k)}(\theta,\varepsilon)$ for ensemble-based phase-reduction analysis.}
\label{fig:PhaseReductionAnalysis}
\end{figure}

For practical implementation of the ensemble-based framework, we consider a finite number of sample cycles to estimate the phase-sensitivity function. We note that sample cycles are extracted based on the observed probabilistic distribution of their occurrence in three-dimensional flows to avoid biased sampling. In this study, since the amplitude variable, $r$, in the phase plane captures the three-dimensionality of the cylinder wake, we construct the probability density function of this cycle amplitude using a kernel density estimation with the Gaussian function \citep{bowman1997applied} as shown in figure~\ref{fig:Sampling}(a). We then divide $P(r)$ into subdivisions that have an equal probability, and sample cycles are extracted from each subdivision. We then apply the direct method to obtain the phase-response functions on these selected samples. Extracted sample cycles are impulsively perturbed at various phases to measure the phase-response function as shown in figure~\ref{fig:PhaseReductionAnalysis}$(a)$. Finally, the phase-sensitivity function is calculated by averaging and normalizing with the impulse magnitude, $\varepsilon$, yielding
\begin{equation}    \label{eq:PhaseSensitivityFunction_EnsembleAveraged}
    \zeta(\theta) \approx \frac{1}{\varepsilon K} \sum_{k=1}^{K} g^{(k)}(\theta;\varepsilon),
\end{equation}
where $g^{(k)}(\theta;\varepsilon)$ denotes the phase-response function \revJK{of} the $k$-th sample cycle from among a total of $K$ cycles.

Determination of the phase-response function poses another challenge in the phase-reduction analysis of three-dimensional flows. Unlike for two-dimensional flows, the perturbed three-dimensional flow does not exactly return to the unperturbed trajectory due to its chaotic nature negating the concept of an asymptotic phase shift. This requires careful assessment of the phase-response function. We observe that when the perturbation is sufficiently weak, there exists a finite time interval where the dynamics of the perturbed flow remain analogous to the unperturbed flow. Within this finite time interval, the perturbed flow initially deviates from the unperturbed flow due to the transient effect of the perturbation. The perturbed flow then approaches the unperturbed flow on the phase plane as the transient effect of the perturbation dies out. After this time horizon, the perturbed flow starts to deviate again from the unperturbed flow due to the chaotic nature of three-dimensional flows.

Based on this perspective, we use the time $t^{*}$, when the perturbed flow \revJK{is} closest to the unperturbed flow, to estimate the phase-response functions of three-dimensional flows. To determine $t^{*}$, we monitor the distance function $d$ between the unperturbed and perturbed flows on the phase plane as shown in figure~\ref{fig:PhaseReductionAnalysis}$(b)$, which is defined as
\begin{equation}
\begin{split}
    d^{(k)}(t;\theta,\varepsilon) =  \Biggl[ & \left(R(\bs{q}_{\delta}^{(k)})\right)^2 + \left(R(\bs{q}_{0}^{(k)})\right)^2 \\
                                    & - 2R(\bs{q}_{\delta}^{(k)})R(\bs{q}_{0}^{(k)})\cos\left(\Theta(\bs{q}_{\delta}^{(k)}) - \Theta(\bs{q}_{0}^{(k)})\right)\Biggr]^{1/2}
\end{split}
\end{equation}
using the law of cosines. \revYJ{Similar to the way we denote the phase functional $\Theta$ as returning the phase $\theta(t) = \Theta(\bs{q}(\bs{x},t))$ for a particular flow $\bs{q}(\bs{x},t)$, the amplitude }functional $R$ returns the amplitude of that flow\revYJ{ field as $r(t) = R(\bs{q}(\bs{x},t))$}. We now seek the time $t^{*}$ such that $d$ is minimized \revJK{after the impulse perturbation and} before it starts to grow again. To not be affected by the local fluctuations of $d$, we mainly track the variation of local minimum and maximum values of $d$. Both local maximum and minimum values of the distance function $d$ keep decreasing as the transient effects of the perturbation diminish. When either the local minimum or maximum value of $d$ increases, we can presume that the perturbed flow \revJK{has started} to deviate from the unperturbed flow. Thus, we take the time $t^{*}$ \revJK{to be} when we observe the minimum value in the function $d$ before either of the extrema values start to increase. Finally, the phase-response function is measured as $g^{(k)}(\theta;\varepsilon) = \Theta(\bs{q}_{\delta}^{(k)}(\bs{x},t^{*})) - \Theta(\bs{q}_{0}^{(k)}(\bs{x},t^{*}))$ for each sample cycle, as shown in figure~\ref{fig:PhaseReductionAnalysis}$(b)$.

In addition to the phase-response and phase-sensitivity functions, we can \revJK{similarly} measure the amplitude variation introduced by the weak impulsive perturbation. Since $r$ denotes the three-dimensional characteristics of the wake, the deviation of the \revJK{ampliutde in the phase plane} trajectory at $t^*$ represents a modification of underlying three-dimensional characteristics. In this context, we also quantify the function $h^{(k)}(\theta;\varepsilon) = R(\bs{q}_{\delta}^{(k)}(\bs{x},t^{*})) - R(\bs{q}_{0}^{(k)}(\bs{x},t^{*}))$ in addition to the phase-response functions as shown in figure~\ref{fig:PhaseReductionAnalysis}$(b)$. Similar to the phase-sensitivity function $\zeta(\theta)$, the function $h$ is averaged and normalized with the impulse magnitude, $\varepsilon$, to give the function $\xi(\theta)$ defined as
\begin{equation}
    \xi(\theta) \equiv \frac{1}{\varepsilon K} \sum_{k=1}^{K} h^{(k)}(\theta;\varepsilon).
\end{equation}
This function represents the mean normalized variation of $r$ made by the impulsive perturbations.

\section{Synchronization characteristics of three-dimensional wake}    \label{sec:Results}

\begin{figure}
    \centering
    \includegraphics[width=\textwidth]{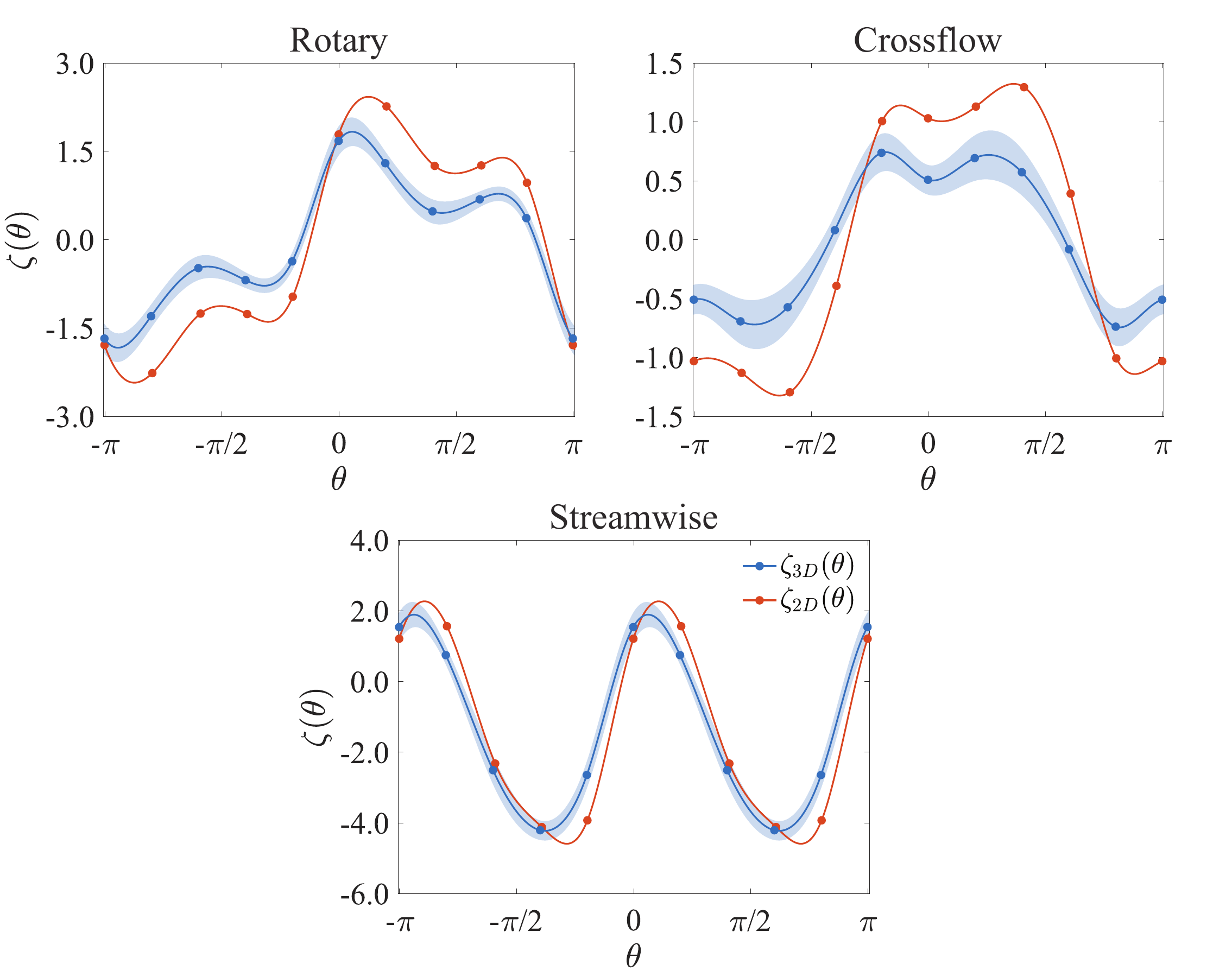}
    \captionsetup{format=plain,justification=justified}
    \caption{Comparison of phase-sensitivity functions of two- and three-dimensional wakes, denoted as $\zeta_{2D}(\theta)$ and $\zeta_{3D}(\theta)$ respectively, at $\Rey =300$ for each perturbation type. The shaded area displays the 95\% confidence interval of $\zeta_{3D}(\theta)$.}
    \label{fig:PhaseSensitivityFunction}
\end{figure}

Let us numerically investigate the synchronization between oscillations of a circular cylinder and its wake based on phase-reduction analysis. In order to study the influence of the three-dimensionality in the wake, we compare the three-dimensional \revJK{wake} synchronization characteristics with the imposed two-dimensional wake \revJK{ones} at the same Reynolds number. As mentioned in \S\ref{sec:Introduction}, we consider three types of in-plane cylinder \revJK{motions}, streamwise translation, crossflow translation, and rotation. \revYJ{While the two-dimensional wake is analyzed as presented in \S\ref{sec:PhaseReductionAnalysis_Periodic}, the three-dimensional wake is studied using the ensemble-based phase-reduction analysis with 10 sample cycles as shown in figure~\ref{fig:Sampling}$(b)$. The sample cycles are extracted based on the probability density function constructed with 200 shedding cycles provided in figure~\ref{fig:Sampling}$(a)$, which is identified to be of sufficient length for the convergence of the cumulative distribution function. We apply the impulsive cylinder motion with a magnitude of $\varepsilon/U_{\infty} = 0.025$, which is identified as being sufficiently small to satisfy the linearity assumption, to measure the phase-sensitivity functions. This identification is performed by testing successively smaller impulse magnitudes and noting that corresponding decreases occurred in the phase or amplitude difference between the perturbed and unperturbed flows. However, decreasing the impulse magnitude from 0.025 to 0.0125 does not produce meaningful changes in the phase or amplitude difference for either of the three types of motions used.}

\revJK{Two- and three-dimensional wake} phase-sensitivity functions for rotary, crossflow \revJK{translation} and streamwise \revJK{translation} motions are compared in figure~\ref{fig:PhaseSensitivityFunction}. The ensemble-averaged $\zeta(\theta)$ for three-dimensional flows are shown in blue and the light blue regions show the 95\% confidence intervals obtained from the samples. We note that the overall trend of \revJK{three-dimensional wake} phase-sensitivity functions resembles those from two-dimensional wakes. This is because the two-dimensional coherent structures in von K\'{a}rm\'{a}n vortex shedding are predominant, even in three-dimensional wakes at $\Rey=300$, over the secondary instabilities. Hence, the perturbation dynamics of the flow follow the shedding of spanwise vortices. This implies that the potential exists to use two-dimensional flows from low to moderate Reynolds numbers to capture phase-sensitivity functions and perturbation dynamics for three-dimensional flows at the equivalent Reynolds number.

\begin{figure}
    \centering
    \includegraphics[width=\textwidth]{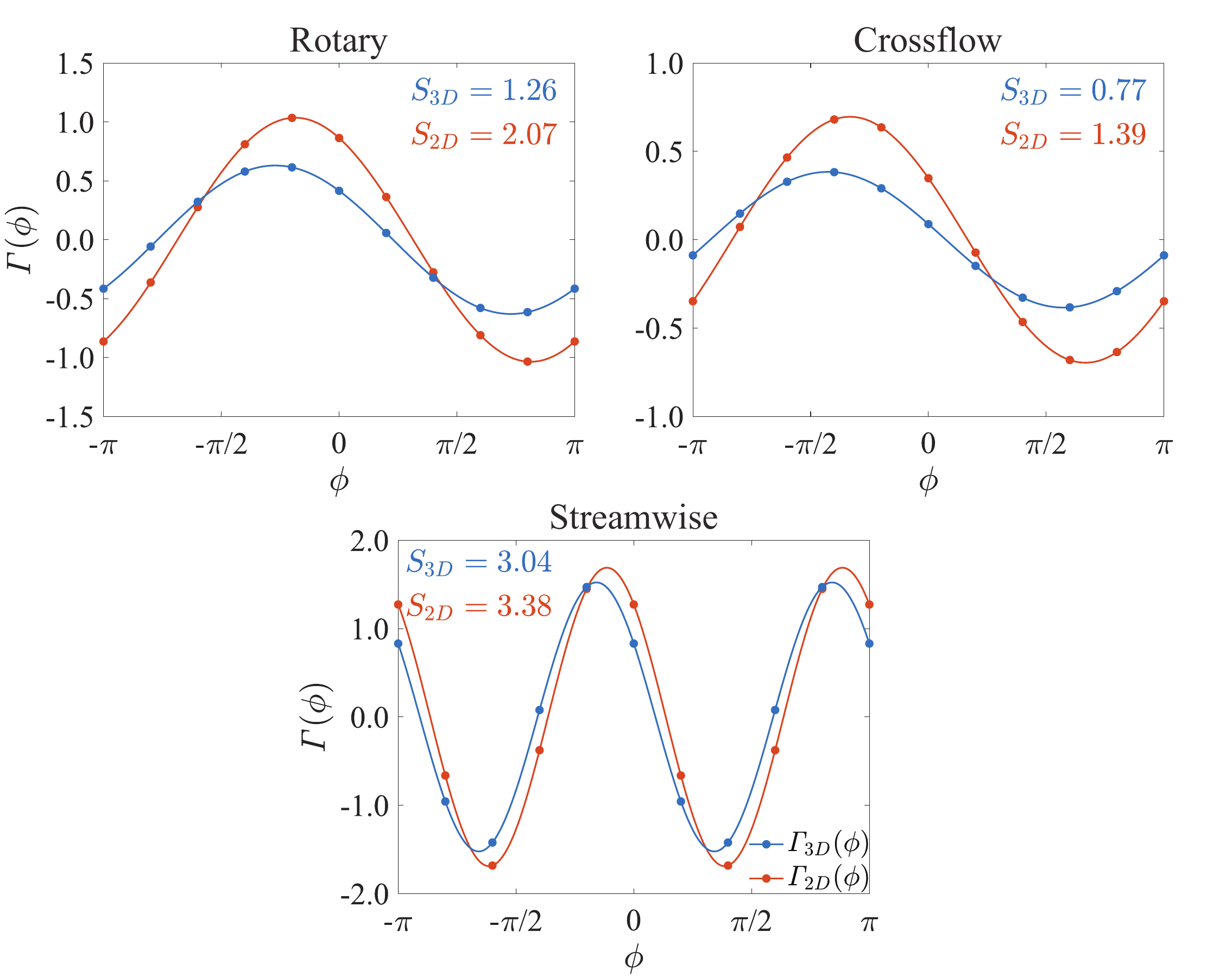}
    \captionsetup{format=plain,justification=justified}
    \caption{Phase-coupling functions, $\mathit{\Gamma}_{2D}(\phi)$ and $\mathit{\Gamma}_{3D}(\phi)$, and synchronizability, $S_{2D}$ and $S_{3D}$, evaluation of two- and three-dimensional wakes for each perturbation type.}
    \label{fig:PhaseCouplingFunction}
\end{figure}

We find that the three-dimensional wake is less sensitive than the two-dimensional wake to all \revJK{three} types of external perturbation in terms of the phase. The magnitudes of \revJK{the} phase-sensitivity functions \revJK{for a} three-dimensional wake is significantly decreased for the rotary and crossflow perturbations. In contrast, the phase-sensitivity function for the streamwise perturbation is much less affected by the three-dimensionality. When the cylinder translates in the streamwise direction, \revJK{it} moves toward or away from the vortices behind it. Thus, a phase shift by the streamwise motion is largely associated with the adjustment of distance between the cylinder and the spanwise vortices. This distance is less affected by the spanwise three-dimensionality of the wake.

Using the phase-sensitivity functions, we determine the synchronizability of the wake to \revJK{steady} sinusoidal oscillations using the phase-coupling functions, $\mathit{\Gamma}(\phi)$. Phase-coupling functions of two- and three-dimensional wakes are evaluated and shown in figure \ref{fig:PhaseCouplingFunction}. Since the synchronization of the cylinder wake to streamwise oscillations mainly occurs when $\Omega_{f} \approx 2\Omega_{n}$ \citep{al2007forced,marzouk2009reduction}, the second harmonic ($m = 2$) is investigated for the streamwise translation. In fact, the phase coupling function with $m = 1$ for streamwise oscillations is zero valued ($\mathit{\Gamma}(\phi) = 0$) due to the $\pi$ periodicity of the phase-sensitivity function as shown in figure~\ref{fig:PhaseSensitivityFunction}, indicating that synchronization to the first harmonic is hardly seen. \revYJ{However, synchronization to the first harmonic of streamwise cylinder oscillations can occur for large enough oscillation amplitudes \citep{leontini2013wake,konstantinidis2016vortex}.}

\begin{figure}
    \centering
    \includegraphics[width=\textwidth]{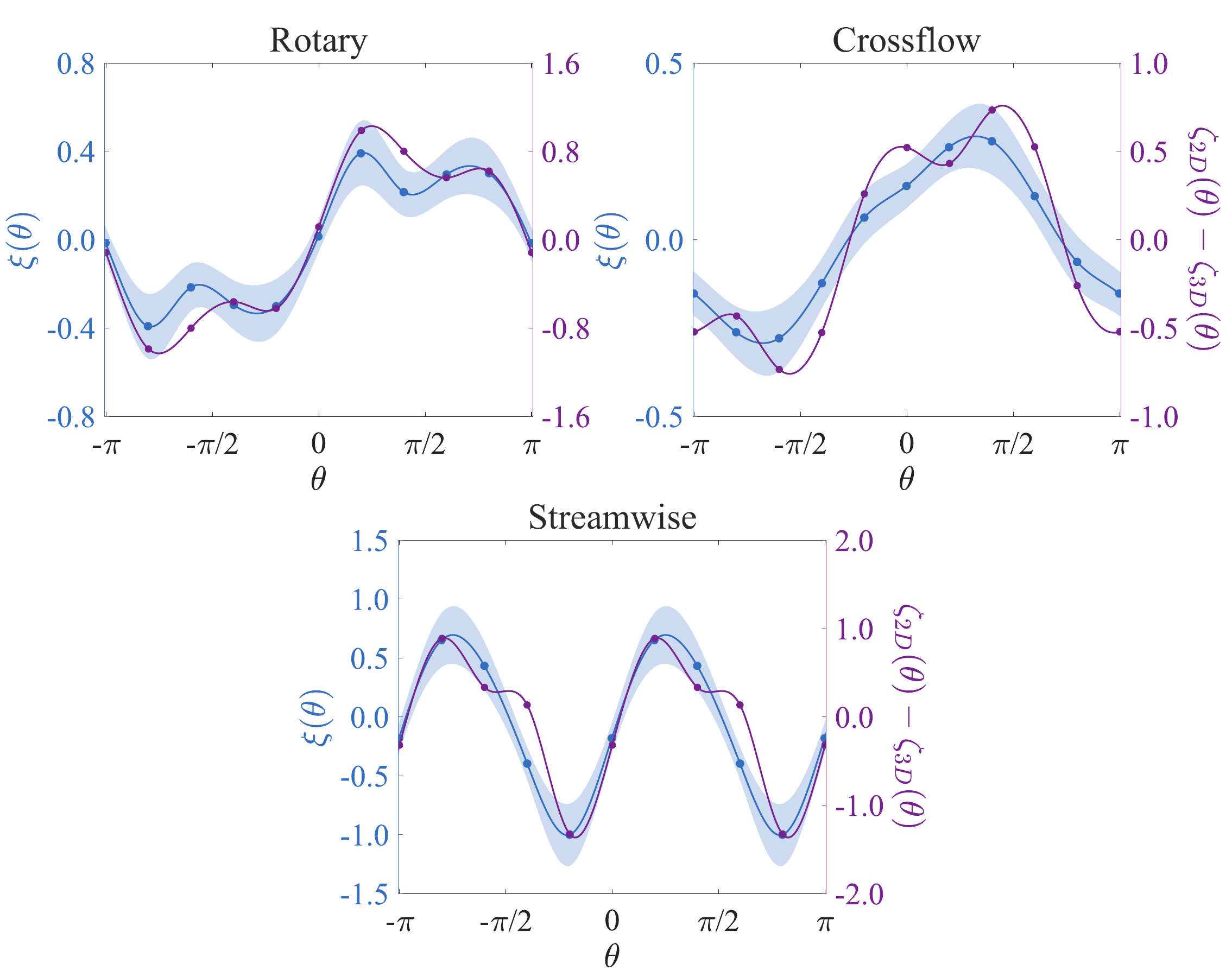}
    \put(-425,335){$\rho=0.977$} \put(-201,335){$\rho=0.954$} \put(-313,153.5){$\rho=0.934$}
    \captionsetup{format=plain,justification=justified}
    \caption{Amplitude variation, \revJK{$\xi(\theta)$, for} three-dimensional \revJK{wakes} \revYJ{normalized }by \revYJ{the magnitude of }impulsive cylinder motion \revJK{along} with 95\% confidence intervals. Differences between the phase-sensitivity functions of two- and three-dimensional wakes, $\zeta_{2D}(\theta) - \zeta_{3D}(\theta)$, are also provided for comparison.}\label{fig:AmplitudeSensitivityFunction}
\end{figure}

On the other hand, the first harmonic ($m = 1$) is targeted for rotation and crossflow translation. Similar to the phase-sensitivity function results, we again observe that three-dimensionality within the \revJK{wake} imposes a higher resistance \revJK{to} synchronization. Synchronizability to rotary and crossflow oscillations shows a large reduction (39.1\% and 44.6\%, respectively) compared to the two-dimensional flow. The synchronizability to streamwise oscillation showed \revJK{only} a 10.1\% decrease.

\begin{figure}
    \centering
    \includegraphics[width=\textwidth]{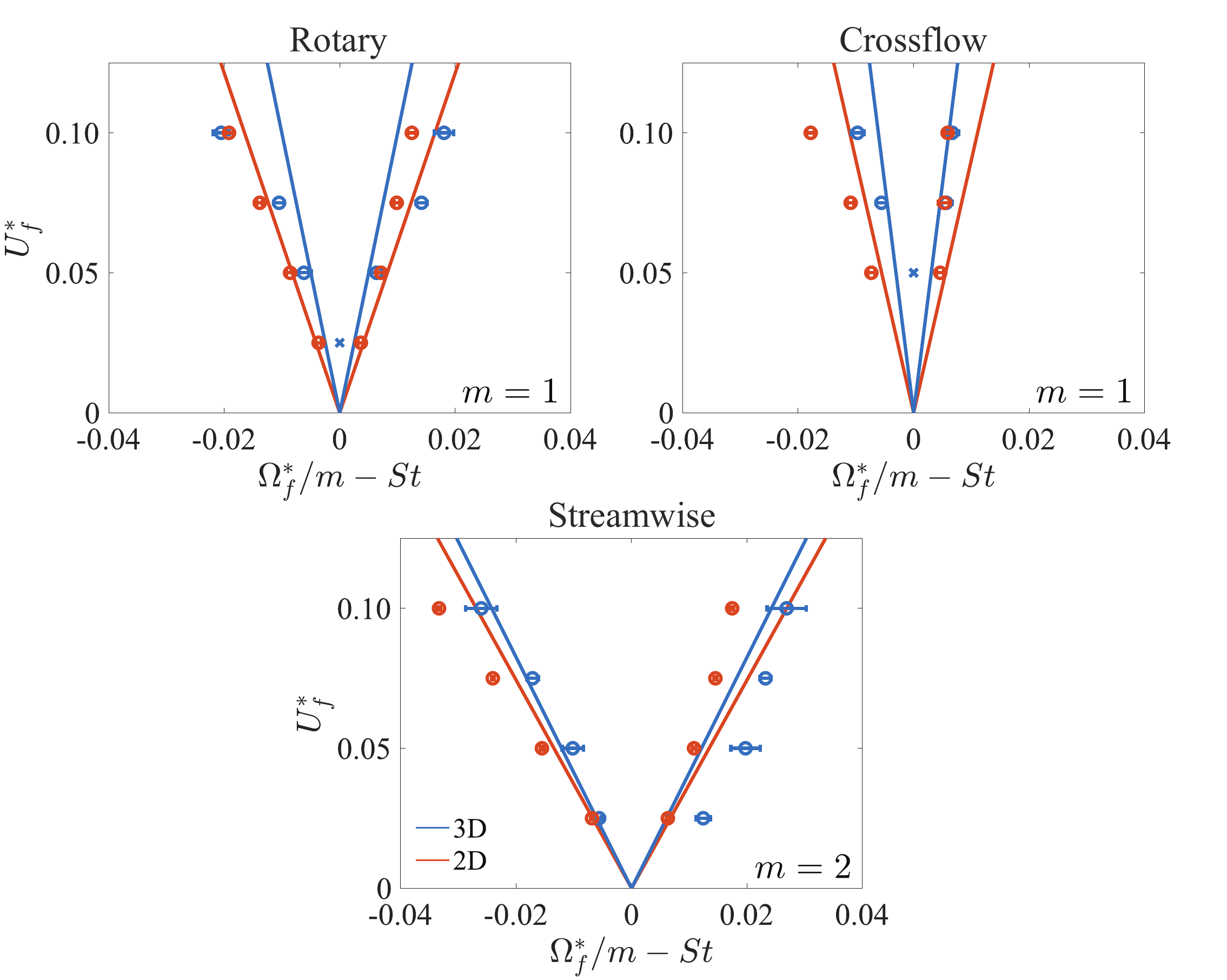}
    \captionsetup{format=plain,justification=justified}
    \caption{Synchronization condition (Arnold tongue) for each type of oscillation \revJK{motion} predicted using phase-reduction analysis (solid lines) and identified using DNS (circles). \revYJ{Horizontal bars represent the uncertainty in the identification of synchronization boundaries by DNS.}}
    \label{fig:LockInBoundary}
\end{figure}

To understand the source of the difference between the phase-sensitivity functions of two- and three-dimensional \revJK{wakes} at the same Reynolds number, we consider the effect of impulsive cylinder motion in an additional degree of freedom of the three-dimensional wake, the amplitude $r$. We compare $\xi(\theta)$ with the difference in phase sensitivities, $\zeta_{2D}(\theta) - \zeta_{3D}(\theta)$, for the three perturbation \revJK{motions} as shown in figure~\ref{fig:AmplitudeSensitivityFunction}. We observe that the function $\xi(\theta)$ generally becomes larger at phases where the phase-sensitivity function of the three-dimensional wake exhibits a relatively large difference from the \revJK{corresponding} two-dimensional \revJK{one}. Correlation coefficients, $\rho$, between $\xi(\theta)$ and $\zeta_{2D}(\theta) - \zeta_{3D}(\theta)$ are also evaluated \revJK{and have values} higher than 0.9. This indicates that the energy given by the cylinder motion changes both the instantaneous phase and the amplitude in the phase plane. Recalling that the amplitude variable $r$ reflects the instantaneous three-dimensional properties of the wake, it can be argued that \revJK{the portion of energy transferred, in the amplitude direction, from the cylinder to the wake modifies} the three-dimensional characteristics of the flow. \revJK{However, since only the energy transferred in the instantaneous phase direction supports synchronization}, three-dimensional wakes require larger energy to achieve synchronization as observed in the synchronizability estimations.

Based on equation~\ref{eq:SynchronizationCondition}, the synchronization conditions of the nondimensional forcing amplitude, $U^{*}_{f} = U_f/U_\infty$, and frequency, $\Omega^{*}_{f}=\Omega_{f}D/(2\pi U_{\infty})$, can be theoretically predicted based on the phase-coupling functions and \revJK{shown} in figure~\ref{fig:LockInBoundary}. We validate our methodology and results by comparing the predicted synchronization boundaries from phase-reduction analysis to DNS parametric studies. The region between the solid lines is predicted to be synchronized to the external forcing while the two regions outside the lines are predicted to be asynchronous. The circles indicate the synchronization bounds determined by performing DNS with the cylinder \revJK{undergoing steady and continuous oscillations} at the corresponding forcing amplitude and frequency. \revYJ{Horizontal error bars represent the uncertainty in the identification of synchronization boundaries associated with the nonuniform and finite resolution of the DNS parametric sweeps.} Synchronization in DNS is determined by monitoring the temporal variation of the phase difference, $\phi(t)$, between the sinusoidally forced cylinder oscillation and the wake, defined in equation~\ref{eq:PhaseDifference}. When $\phi(t)$ converges to a certain value with a small standard deviation, $\sigma(\phi(t)) < 0.1\pi$, we classify the case as synchronized. 

From figure~\ref{fig:LockInBoundary}, we observe that phase-reduction analysis is able to predict the synchronization conditions for both two- and three-dimensional wakes for $U^{*}_{f} < 0.1$, indicating that the linear assumption is valid for oscillations with small amplitudes. As the oscillation amplitude of the nondimensional velocity increases, the DNS results deviate from the predictions due to the stronger nonlinearity. In addition, synchronization to rotary and crossflow oscillations \revJK{are} not observed when the oscillation amplitude is weak, classified as $U^{*}_{f} = 0.025$ and $0.05$ \revJK{for each type of motion}, respectively, \revJK{and denoted as cross symbols in the figure}. As discussed in \citet{pikovsky1997phase}, chaotic oscillators with phase diffusion do not show synchronization when the oscillation is not strong enough to suppress the phase diffusion. 

To elaborate on this point theoretically, we look into the temporal variation of the instantaneous phase difference $\phi(t)$, which can be approximately expressed as
\begin{equation}
    \dot{\phi}(t) \simeq \left(\Omega_{n} - \frac{\Omega_{f}}{m}\right) + F(\bs{q}(\bs{x},t)) + U^{*}_{f}\eta(t)\zeta\left(\phi + \frac{\Omega_{f}}{m} t \right)
\end{equation}
derived from equation~\ref{eq:PhaseDynamics_wPerturbation_PseudoPeriodic}. Let us now compare the scales of each term on the right-hand side of this equation under the \revJK{assumption} that the oscillation amplitude, $U^{*}_{f}$, is small and approaches zero. Since $\Omega_{f}/m$ should be very close to $\Omega_{n}$ to synchronize the wake to small-amplitude oscillations, the first term is negligible. The third term is also negligible because of the linear proportionality to $U^{*}_{f}$. \revJK{However}, the second term $F$ is independent of the external forcing because it is an inherent property of the three-dimensional wake. Thus, the first and third terms disappear for small-amplitude oscillation conditions, and the dynamics of $\phi$ \revJK{are} governed predominantly by $F$ as $\dot{\phi}(t) \simeq F(\bs{q}(\bs{x},t))$. As a result, if instantaneous fluctuations $F$ in three-dimensional flows are not sufficiently small, it prevents the convergence of $\phi$ unless $F$ is restrained by relatively large oscillations. \revJK{Thus, the} oscillation amplitude has to exceed a certain threshold to facilitate synchronization \revYJ{regardless of the forcing frequency}. Furthermore, the third term is also proportional to the phase-sensitivity function, which implies a larger threshold for $U^{*}_{f}$ when the magnitude of the phase-sensitivity function $\zeta(\theta)$ is smaller. This is seen in the largest threshold for $U^{*}_{f}$ for the crossflow oscillation and no threshold appearing within $U^{*}_f \geq 0.025$ for the streamwise oscillation, which correspond to the smallest and largest phase-sensitivity function \revJK{magnitudes} among oscillation types as shown in figure~\ref{fig:PhaseSensitivityFunction}.

\begin{figure}
    \centering
    \includegraphics[width=\textwidth]{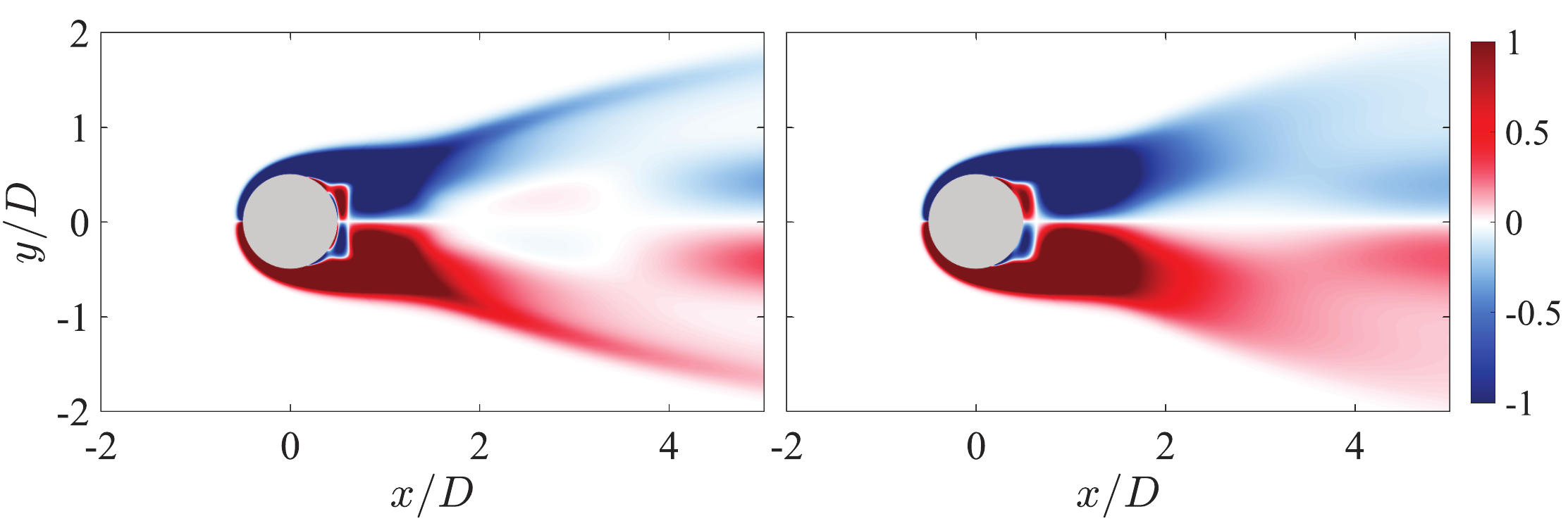}
    \captionsetup{format=plain,justification=justified}
    \caption{Contours of mean spanwise vorticity of the two-dimensional (left) and three-dimensional (right) wakes.}
    \label{fig:VortexFormationRegion}
\end{figure}

We also note that the left synchronization boundaries obtained from the \revJK{three-dimensional wake} DNS results are located inside the two-dimensional wake \revJK{results}. This indicates the added difficulty in synchronizing the three-dimensional wake as predicted by phase-reduction analysis. On the \revJK{other hand}, the right synchronization boundaries, \revJK{representing} high-frequency oscillations, exhibit different trends depending on the oscillation type which \revJK{deviates from} the theoretical predictions. 

For two-dimensional wakes, we commonly see deviations of the \revJK{DNS identified} right boundaries from the predictions by phase-reduction analysis, which is also observed in the previous study of \citet{khodkar2021phase}. They identified that synchronization to higher frequencies is more restrictive than lower frequencies, attributing it to the physical constraint of the vortex formation region behind the cylinder. Synchronization to high-frequency oscillations requires the vortex formation region to be compact. However, it is impossible to shorten the vortex formation region beyond a certain limit. This is reflected in the asymmetry of the DNS-based Arnold tongue and its inward deflection at the right branch for two-dimensional wakes. \revJK{This hypothesis, however, only applies to } two-dimensional wakes as we observe that three-dimensional wakes exhibit a longer vortex formation region than the \revJK{corresponding} two-dimensional wake, as visualized in figure~\ref{fig:VortexFormationRegion}. Since a three-dimensional wake initially has a longer vortex formation region, it holds a \revJK{greater} capacity for shortening the vortex formation region. This offers an easier path to synchronization for high-frequency oscillations \revJK{in a three-dimensional wake} compared to a two-dimensional wake, resulting in no inward deflection of the Arnold tongue at the right synchronization boundary.

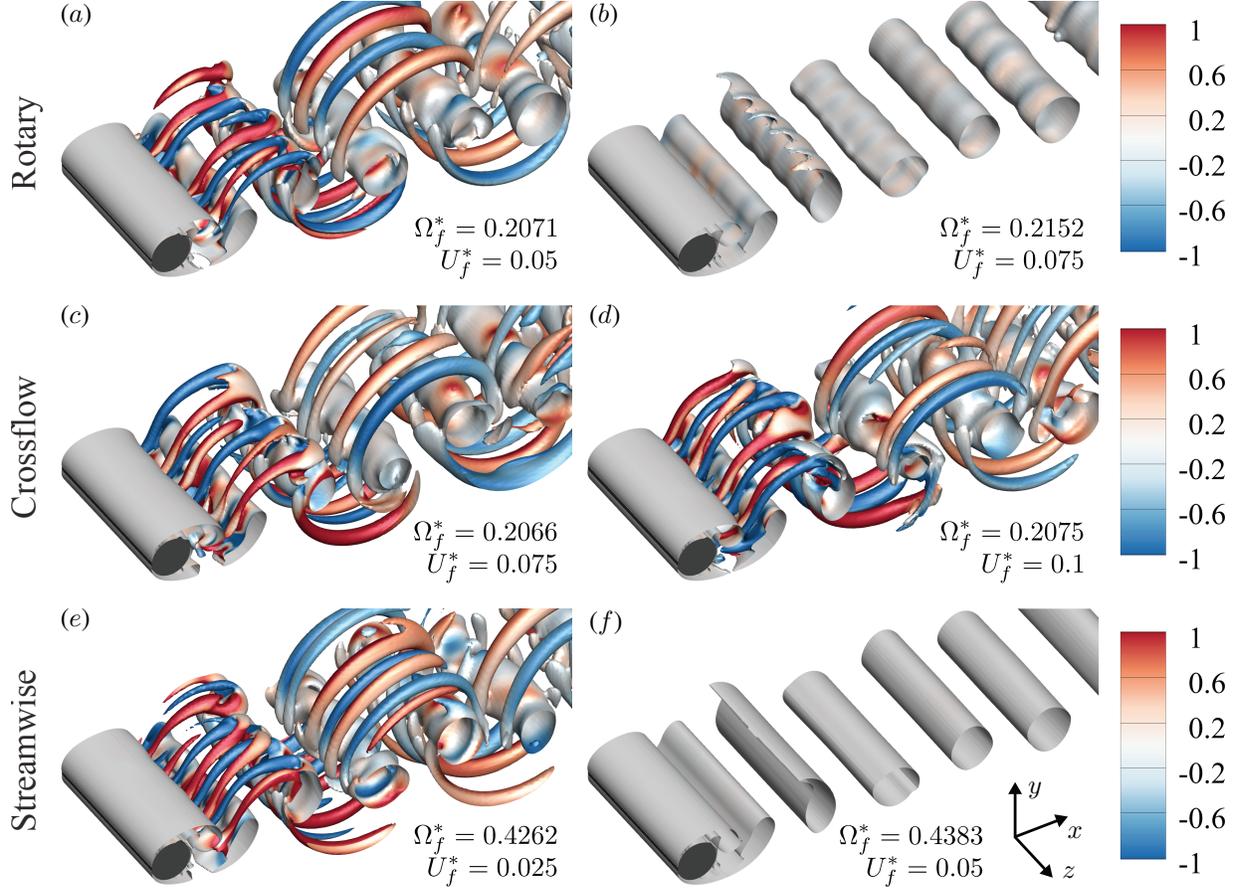
\begin{figure}
    \centering
    \includegraphics[width=\textwidth]{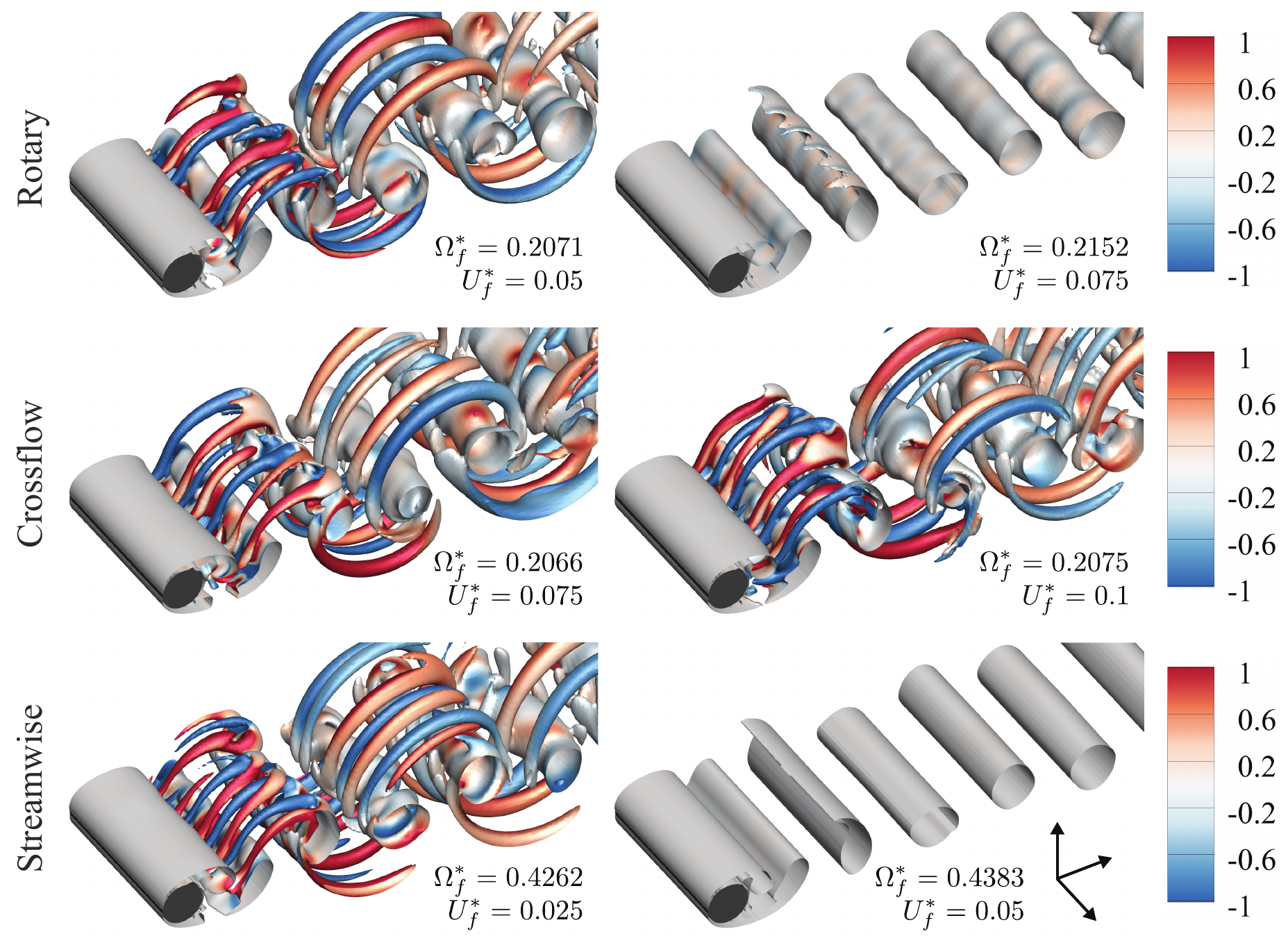}
    \put(-445,332){($a$)}    \put(-445,218){($c$)}    \put(-445,103){($e$)}
    \put(-245,332){($b$)}    \put(-245,218){($d$)}    \put(-245,102){($f$)}
    \put(-64,23){$x$}    \put(-79,40){$y$}    \put(-66,8){$z$} 
    \captionsetup{format=plain,justification=justified}
    \caption{Wake visualization with isosurfaces of the $Q$-criterion ($Q=0.05$) colored by $\omega_{x}$ for \revYJ{cases} on the right synchronization boundaries.}
    \label{fig:FlowStructures_Oscillation}
\end{figure}

For the rotary oscillations, while the left synchronization boundary agrees with the prediction, we find that the right synchronization boundary of the three-dimensional wake obtained from DNS results deflects outward as the oscillation amplitude increases. This results in the three-dimensional wake having a larger upper bound than the two-dimensional wake when $U^{*}_{f} \geq 0.075$. Two instantaneous flow structures of cases on the right boundary with $U^{*}_{f} = 0.05$ and $0.075$ for rotary motion are compared in figure~\ref{fig:FlowStructures_Oscillation}(a) and (b). It is observed that secondary vortical structures disappear when $U^{*}_{f} \geq 0.075$, showing that high-frequency rotary oscillations largely suppress the three-dimensionality of the wake. \revYJ{The suppression of secondary instabilities and the two-dimensionalization of the wake by oscillatory cylinder motion are observed in previous studies as well \citep{poncet2002vanishing,deng2007vanishing,kim2009direct}.} This demonstrates the \revJK{existence of a} transition from a three-dimensional to a two-dimensional \revJK{wake} as the oscillation amplitude increases along the right branch. Due to this transition, cylinder wakes with high-frequency rotary oscillation deviate significantly from the unperturbed three-dimensional wake, which gives rise to the strong nonlinear response. Since phase reduction analysis adopts a first-order approximation to estimate the phase-sensitivity function, \revJK{these} nonlinear effects and the transition from a three-dimensional flow to a two-dimensional flow are not captured and are attributed to higher-order effects. On the other hand, such an outward deflection is not observed at the right boundary \revJK{for} crossflow oscillations since the wake transition does not occur \revJK{for this type of motion} as shown in figure~\ref{fig:FlowStructures_Oscillation} $(c)$ and $(d)$.

\begin{figure}
    \centering
    \includegraphics[width=\textwidth]{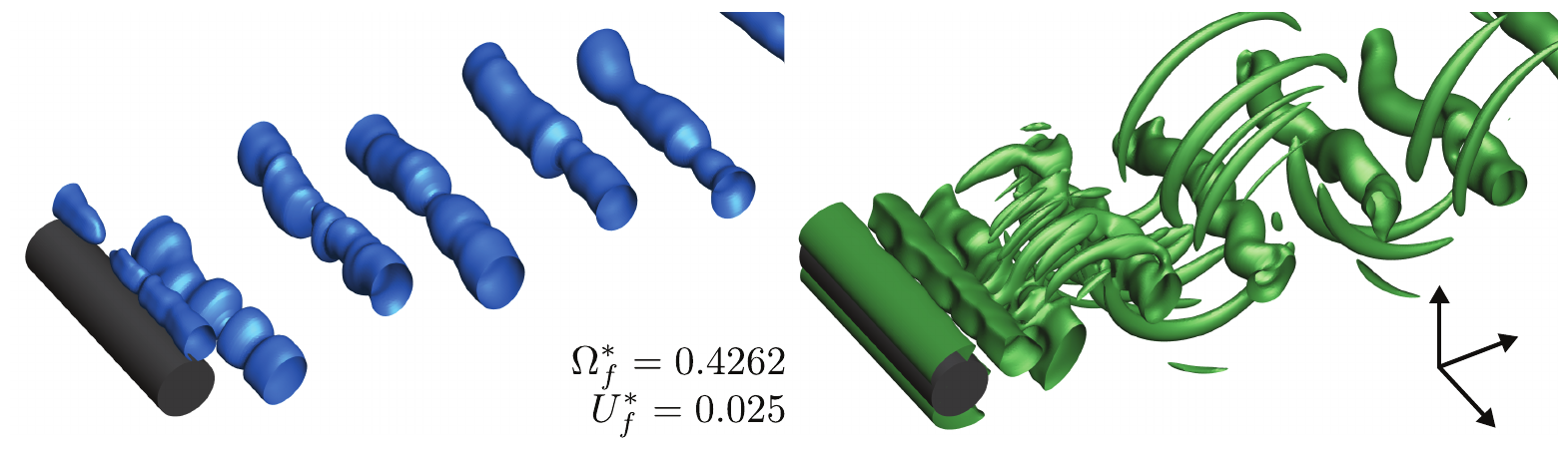}
    \put(-16,24){$x$}    \put(-33,45){$y$}    \put(-19,6){$z$} 
    \captionsetup{format=plain,justification=justified}
    \caption{Scale-decomposed flow structures by isosurfaces of the $Q$-criterion ($Q=0.05$) corresponding to  \revYJ{streamwise oscillation with $U^{*}_{f}=0.025$ and $\Omega^{*}_f=0.4262$ on the right synchronization boundary} . The left and right columns \revJK{capture} large- and small-scale structures, respectively.}
    \label{fig:FlowStructure_ScaleDecomposed_Streamwise}
\end{figure}

Regardless of the oscillation amplitude, the DNS-based right synchronization boundary of the three-dimensional wake is always outside of the two-dimensional wake for the streamwise oscillations. Similar to the rotary oscillations, streamwise oscillations with $U^{*}_{f}=0.05$ \revJK{are able to} two-dimensionalize the cylinder wake, while the three-dimensionality still remains at the oscillation amplitude of $U^{*}_{f}=0.025$, shown in figure~\ref{fig:FlowStructures_Oscillation} (e) and (f). \revJK{To} identify differences \revJK{between} the wakes and understand this \revJK{observation} further, we additionally \revJK{examine} scale-decomposed flow structures for \revJK{this case} on the right boundary, plotted in figure~\ref{fig:FlowStructure_ScaleDecomposed_Streamwise}, and compare them to the unperturbed three-dimensional wakes provided in figure~\ref{fig:FlowStructure_ScaleDecomposed}. We find that both large- and small-scale structures resemble the unperturbed three-dimensional wake with large $r$. This indicates that the three-dimensionality of the wake can be partially suppressed by the streamwise oscillation even though the oscillation amplitude is small. Due to the stronger nonlinear response of the wake to the streamwise motion, discrepancies between the synchronization boundaries of \revJK{the} DNS results and the phase-reduction analysis can be large for this type of oscillation.

\section{Conclusions}    \label{sec:Conclusions}

We examined the influence of the three-dimensionality of the wake on its ability to synchronize to cylinder oscillations at $\Rey = 300$. Phase-reduction analysis was \revJK{used for the} theoretical analysis of wake synchronization. Even though the ability of phase-reduction analysis to study the synchronization of two-dimensional flows has been shown in previous studies, the chaotic nature of three-dimensional wakes makes its implementation for \revJK{these} realistic flows challenging. To address this \revJK{difficulty}, we introduced an ensemble-averaging technique to extend phase-reduction analysis to three-dimensional flows. In addition, we provided a guideline to perform the ensemble-averaged phase-reduction analysis based on the direct method, which includes sample cycle extractions and measurements of phase-sensitivity functions. We applied phase-reduction analysis to both two- and three-dimensional cylinder wakes to understand the effect of three-dimensionality on synchronization.

While the behavior of the phase-sensitivity functions for two- and three-dimensional wakes are similar, due to two-dimensional von K\'arm\'an vortex shedding \revJK{being the dominant feature} in the wake dynamics \revJK{of both}, there is still a significant decrease in the phase-sensitivity of three-dimensional wake. This decrease is especially large for the crossflow and rotary cylinder motions. This \revJK{ultimately} leads to reduced synchronizability and results in narrower Arnold tongues compared to the two-dimensional wake \revJK{cases}. Observing the amplitude variations from impulsive motions, the reduction of synchronizability was attributed to the energy transferred to the wake \revJK{partially going into} the amplitude direction. We identified that the synchronization conditions derived by phase-reduction analysis are valid for cylinder oscillations \revJK{with velocities} that are smaller than 10\% of the free-stream velocity. \revJK{Additionally,} the parametric examination showed that there exists a \revJK{minimum required} oscillation amplitude to overcome the phase diffusion of the three-dimensional wake and achieve synchronization. Nonlinear effects from wake three-dimensionality, which are associated with the vortex formation and transition from three- to two-dimensional flow, are also identified by analyzing flow structures of synchronized wakes. To sum up, the phase-reduction analysis predicts the synchronization conditions with reasonable accuracy, especially for small oscillation amplitudes when nonlinear effects are fairly small.

The present study not only reveals synchronization conditions of the three-dimensional cylinder wake but also provides a systematic way to predict and analyze the synchronization characteristics of more complex flows. \revYJ{Previous experimental studies observed that turbulent cylinder wakes at high Reynolds numbers (up to $\Rey \thicksim 10^5$) also exhibit rhythmic behaviors with a single dominant frequency, according to their energy spectrum, similar to the low Reynolds number wakes \citep{szepessy1992aspect,braza2006turbulence}. Therefore, it is anticipated that phase characterization of the flow is still valid, and the ensemble-based phase-reduction framework would be applicable, to turbulent flows with a high Reynolds number.} Considering the sizable computational costs or experimental resources required to perform parametric studies, phase-reduction analysis gives an effective prediction of the synchronization with reasonable precision.

\section*{Acknowledgments}{This work was supported by the National Science Foundation (NPS Award: 2129638; UCLA Award: 2129639). This work used Expanse at the San Diego Supercomputing Center (SDSC) through allocation PHY230016 from the Advanced Cyberinfrastructure Coordination Ecosystem: Services \& Support (ACCESS) program and Hoffman2 provided by UCLA Office of Advanced Research Computing’s Research Technology Group. \revYJ{The authors are very grateful to Dr. Kai Fukami for his insight and suggestions during this work.}\label{sec:acknowledgments}
}

\section*{Declaration of interests}{

The authors report no conflict of interest.

}

\bibliographystyle{unsrtnat}
\bibliography{reference}

\end{document}